\documentclass[10pt,preprint]{aastex}
\usepackage{graphicx}
\usepackage{natbib}
\usepackage{amssymb}

\newcommand{\s}{\ensuremath{\mathrm{sec}}}

\newcommand{\gcc}{\ensuremath{\mathrm{g\ cm^{-3} }}}
\newcommand{\erg}{\ensuremath{\mathrm{erg}}}
\newcommand{\cm}{\ensuremath{\mathrm{cm}}}
\newcommand{\metre}{\ensuremath{\mathrm{m}}}
\newcommand{\km}{\ensuremath{\mathrm{km}}}

\newcommand{\cms}{\ensuremath{\mathrm{cm \ s^{-1}}}}
\newcommand{\kms}{\ensuremath{\mathrm{km \ s^{-1}}}}

\newcommand{\MeV}{\ensuremath{\mathrm{MeV}}}
\newcommand{\kelvin}{\ensuremath{\mathrm{K}}}

\newcommand{\element}[2]{\ensuremath{\mathrm{{}^{#2}{#1}}}}
\newcommand{\massfrac}[2]{\ensuremath{X_\element{#1}{#2}}}
\newcommand{\cfrac}{\massfrac{C}{12}}
\newcommand{\ofrac}{\massfrac{O}{16}}
\newcommand{\netwentyfrac}{\massfrac{Ne}{20}}
\newcommand{\netwentytwofrac}{\massfrac{Ne}{22}}
\newcommand{\partf}[2]{\ensuremath{\frac{\partial #1}{\partial #2}}}
\newcommand{\drvf}[2]{\ensuremath{\frac{d #1}{d #2}}}
\newcommand{\FLASH}{{\sc{Flash}}{}}
\newcommand{\eg}{{\it{e.g.}}}
\newcommand{\kmax}{\ensuremath{\kappa_{\mathrm{max}}}}
\newcommand\citeeg[1]{\citep[\eg{},][]{#1}}

\newlength{\curcolwidth}   

\begin{document}

\setlength{\curcolwidth}{\columnwidth} 

\title{Local Ignition in Carbon/Oxygen White Dwarfs -- I: One-zone Ignition and Spherical Shock Ignition of Detonations}
\shorttitle{Local Ignition and Shock-Ignited Detonations}
\shortauthors{Dursi \& Timmes}

\author{L. Jonathan Dursi}
\affil{Canadian Institute for Theoretical Astrophysics,
       University of Toronto, 
       60 St. George St.,
       Toronto, ON, M5S~3H8, 
       Canada}
\email{ljdursi@cita.utoronto.ca}
\author{F. X. Timmes}
\affil{Theoretical Division, 
       Los Alamos National Laboratory, 
       Los Alamos, NM, 87545, 
       USA}
\email{timmes@lanl.gov}

\begin{abstract}
The details of ignition of Type Ia supernovae remain fuzzy, despite the
importance of this input for any large-scale model of the final explosion.
Here, we begin a process of understanding the ignition of these hotspots
by examining the burning of one zone of material, and then investigate
the ignition of a detonation due to rapid heating at single point.

We numerically measure the ignition delay time for onset of burning in
mixtures of degenerate material and provide fitting formula for conditions
of relevance in the Type~Ia problem.  Using the neon abundance as a
proxy for the white dwarf progenitor's metallicity, we then find that
ignition times can decrease by $\sim 20\%$ with addition of even 5\%
of neon by mass.  When temperature fluctuations that successfully kindle
a region are very rare, such a reduction in ignition time can increase
the probability of ignition by orders of magnitude.  If the neon comes
largely at the expense of carbon, a similar increase in the ignition time
can occur.

We then consider the ignition of a detonation by an explosive energy input
in one localized zone, \eg{} a Sedov blast wave leading to a shock-ignited
detonation.  Building on previous work on curved detonations, we confirm
that surprisingly large inputs of energy are required to successfully
launch a detonation, leading to required matchheads of $\approx$ 4500
detonation thicknesses -- tens of centimeters to hundreds of meters --
which is orders of magnitude larger than naive considerations might
suggest.  This is a very difficult constraint to meet for some pictures
of a deflagration-to-detonation transition, such as a Zel'dovich gradient
mechanism ignition in the distributed burning regime.  
\end{abstract}


\section{INTRODUCTION}
\label{sec:intro}

The current favored model for Type Ia supernovae (SNIa) involves burning
beginning as a subsonic deflagration near the central region of a
Chandrasekhar-mass white dwarf.  Progress has been made in recent years
in understanding the middle stages of these events through multiscale
reactive flow simulations where the initial burning is prescribed as
an initial condition of one or more sizable bubbles already burning
material at time zero.  However, the initial ignition process by
which such bubbles begin burning -- whether enormous 50~km bubbles
\citep{pcl} or more physically motivated smaller igniting points
\citep{mpa02,hoeflichstein02,barcelona03} remains poorly understood.
Further, if later in the evolution there is a transition to a detonation
\citep[\eg{},][]{nrl04}, this ignition process, too, must be explained.
Indeed, ignition physics will play a role --- by determining the
location, number, and sizes of the first burning points --- in any
currently viable SNIa model.  However, until very recently \citep[for
example,][]{woosley04,barcelona05} very little work has gone into
examining the ignition physics of these events.  Here we begin examining
the ignition process by considering the simplest ignitions possible --
that of a single zone -- and the possibility of igniting a detonation
from a Sedov blast wave launched at a single point.

\subsection{Ignition Delay Times}

Astrophysical combustion, like most combustion \citep[for
  example,][]{williams,glassman96}, is highly temperature-dependent;
the $^{12}$C + $^{12}$C reaction, for example, scales as $T^{12}$ near
10$^{9}$ K.  Rates for the exothermic reactions which define the burning
process are generally exponential or near-exponential in temperature
\citep[\eg{},][]{cf88}.  Thus a region with a positive temperature
perturbation can sit `simmering' for a very long time, initially only very
slowly consuming fuel and increasing its temperature as an exponential
runaway occurs.   This is especially true in the electron-degenerate
environment of a white dwarf, where the small increases in temperature
that occur for most of the evolution of the hotspot will have only
extremely small hydrodynamic effects.

If fuel depletion and hydrodynamical effects were ignored, the
temperature of the spot would become infinite after a finite period
of time.  This time is called the ignition time, or ignition delay time,
or sometimes induction time, $\tau_i$.  After ignition starts, burning
proceeds for some length of time $\tau_b$.

\begin{figure}[htb]
\centering
\includegraphics[width=0.45\textwidth]{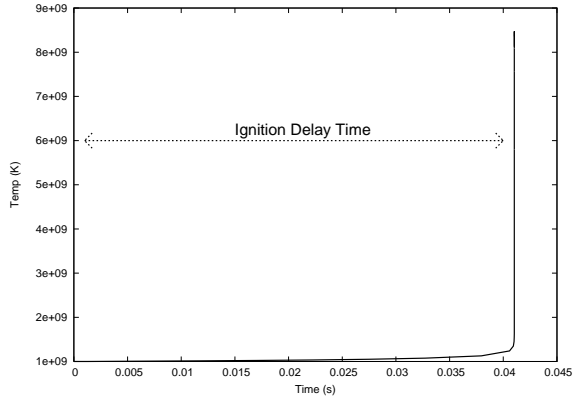}
\caption{ Temperature evolution for burning a zone at constant pressure
with an initial state of $\cfrac = 1.0$, $T = 10^9 \kelvin$, $\rho =
5 \times 10^8 \gcc$.  Because of the strong temperature dependence,
a runaway takes place and most of the burning happens `all at once'.}
\label{fig:ignitiondelay}
\end{figure}

For burning problems of interest, of course, fuel depletion is important,
and no quantities become infinite; however, the idea of an ignition delay
time still holds (see Fig.~\ref{fig:ignitiondelay}).   If the energy
release rate for most of the evolution of the burning is too small to
have significant hydrodynamical effects, and if the timescale over which
burning `suddenly turns on' is much shorter than any other hydrodynamical
or conductive timescales, then the burning of such a hotspot can be
treated, as an excellent approximation, as a step function where all
energy is released from $t = \tau_i$ to $t = \tau_i + \tau_b$.   In many
problems, where $\tau_b \ll \tau_i$, this can be further simplified to
burning occurring only at $t = \tau_i$.   Where such an approximation
(often called `high activation-energy asymptotics') holds, it greatly
simplifies many problems of burning or ignition, reducing the region of
burning in a flame to an infinitesimally thin `flamelet' \citep{mandm}
surface, for instance, or the structure of a detonation to a `square
wave' \citep{erpenbeck}.  Where this approximation does not hold --
such as if slow $\beta$-decay processes are important as bottlenecks
for reactions to proceed (\eg{}, $p$-$p$ burning or the CNO cycle) the
simplification of burning happening only over $\tau_i \le t \le \tau_i +
\tau_b$ often remains useful.

Even for the simple case of one zone, ignition delay times are
relevant for investigation of ignition in SNIa because it sets a
minimum time scale over which an initial local positive temperature
perturbation (hot spot) can successfully ignite and launch a
combustion wave; other timescales, such as turbulent disruption of the
hotspot, or diffusive timescales, must be larger than this for
ignition to successfully occur.

If the burning occurs in an ideal gas, or in a material with some
other simple equation of state, it is fairly easy to write down
approximate ignition times for various burning laws.  In a white
dwarf, however, where the material is partially degenerate or
relativistic and the equation of state is quite complicated
\citep{eos}, no such closed-form expression can be written.  In \S2 we
numerically follow the abundance and thermodynamic evolution of a zone
of white dwarf material in order to measure the ignition times as a
function of the initial temperature, density, and composition.  We
follow both constant-density and constant-pressure trajectories.  The
results are summarized by simple, moderately accurate, fitting
formula.  In \S3 we consider the ignition of a detonation through a
localized energy release producing a Sedov blast wave, and estimate
the amount of energy that must be released for the detonation to
successfully ignite.  In \S4 we consider our results in light of likely
temperature fluctuation spectra during the simmering, convective
phase.

\section{ONE ZONE IGNITION TIMES}
\subsection{Calculations}

We performed a series of 1-zone calculations for the purposes of
measuring ignition times in carbon-oxygen mixtures.  For each of these two
burning conditions -- burning at constant volume and constant pressure
-- over 3500 initial conditions were examined, in a grid of initial
densities, temperatures, and initial carbon fraction.  The carbon (mass)
fractions were in the range $0.4 \le \cfrac \le 1.0$, with the remainder
being oxygen ($\ofrac = 1 - \cfrac$), temperatures of $0.5
\le T_9 \le 7$, and densities $0.1 \le \rho_8 \le 50$, where $T_9$
is the temperature in units of $10^9$~K and $\rho_8$ is the density
in units of $10^8$~g~cm${}^{-3}$.  For burning, a 13-isotope $\alpha$
chain was used \citep{burn}, and a Helmholtz free energy based stellar
equation of state \citep{eos} maintained the thermodynamic state.
The temperature and abundance evolution equations were integrated together
consistently, as described in Appendix~\ref{sec:thermburncoupling}.
Time evolutions were generated as in Fig.~\ref{fig:ignitiondelay}.

To cover the wide range of burning times within each integration, the
timestep was increased or decreased depending on the rate of abundance,
energy release, and thermodynamic changes.  The time interval was varied
by up to a factor of two in each timestep, to try to keep the amount
of energy released through burning per timestep change of fuel within
the range $10^{-5} - 10^{-7}$.  Timesteps failing this criterion were
undone and re-taken with a smaller time interval.   The final ignition
time, when the simulation was stopped, was defined to be time when 90\%
of the carbon was consumed, although the time reported was found to be
insensitive to endpoint chosen.  The code used in this integration, as well as
the resulting data, is available at {\url{http://www.cita.utoronto.ca/$\sim$ljdursi/ignition/}}.

\begin{figure}[htb]
\centering
\plotone{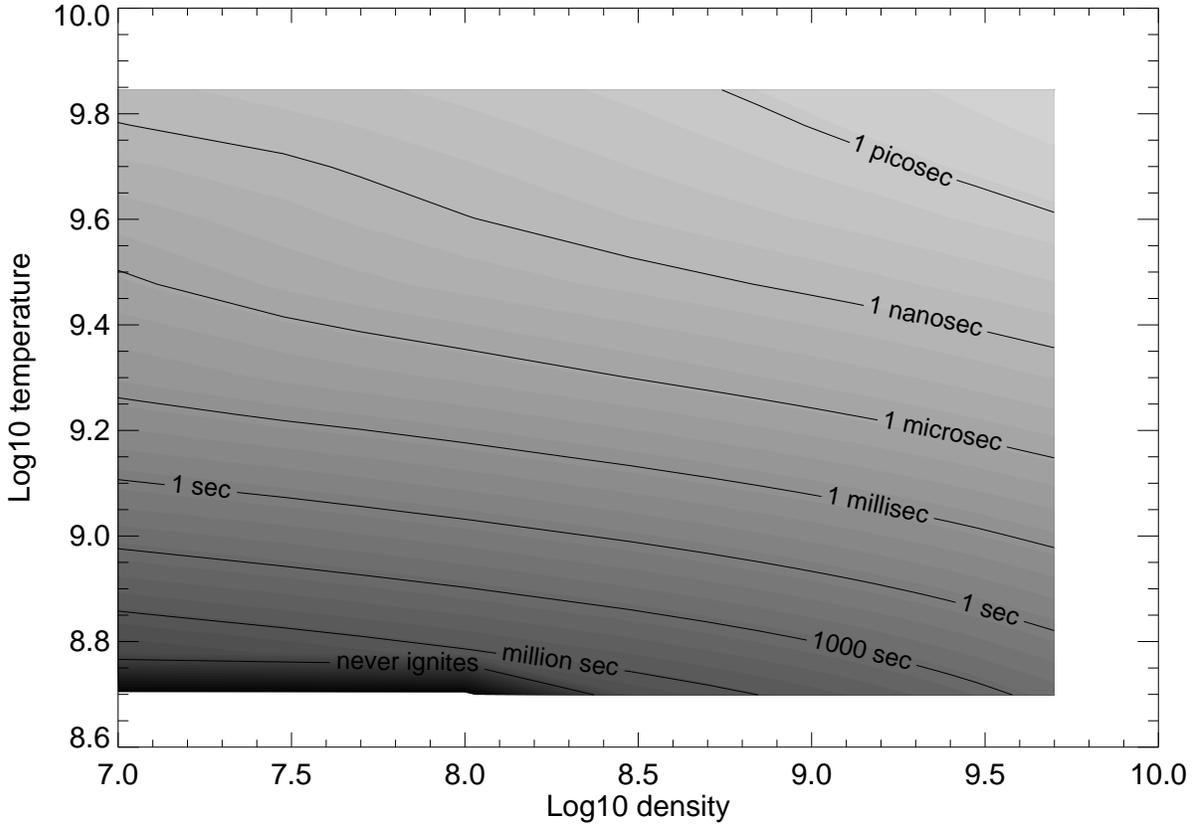}
\caption{ Contour plot of ignition time as a function of initial
density and temperature for a constant-pressure ignition of a mixture
of half-carbon, half-oxygen by mass. }
\label{fig:cp05}
\end{figure}

Over the initial conditions chosen, ignition times varied from
$10^{-14}$~s to $10^{+8}$~s.   A representative contour plot showing the
calculated ignition delay times are shown in Fig.~\ref{fig:cp05}.

Fitting formula for the ignition delay times under the two burning
conditions were determined; the ignition time for constant-pressure
ignition can be given approximately as
\begin{eqnarray}
\tau_{i,{\mathrm{cp}}}(\rho,T,\cfrac) & = & 1.15 \times 10^{-5} \, \s \, \left (\cfrac \rho_8 \right )^{-1.9} f_{\mathrm{cp}}(T) \left ( 1 + 1193 f_{\mathrm{cp}}(T) \right ) \\
f_{\mathrm{cp}}(T) & = & \left( T_9 - 0.214 \right)^{-7.566} \nonumber 
\label{eq:cpfit}
\end{eqnarray}
and that for constant-volume ignition 
\begin{eqnarray}
\tau_{i,{\mathrm{cv}}}(\rho,T,\cfrac) & = & 1.81 \times 10^{-5} \, \s \, \left (\cfrac \rho_8 \right )^{-1.85} f_{\mathrm{cv}}(T) \left ( 1 + 1178 f_{\mathrm{cv}}(T) \right ) \\
f_{\mathrm{cv}}(T) & = & \left( T_9 - 0.206 \right)^{-7.700} \nonumber 
\label{eq:cvfit}
\end{eqnarray}
where $T_9$ is the initial temperature in units of $10^9 \kelvin$, and $\rho_8$ is the initial density in units of 
$10^8 \gcc$.

\begin{figure}[htb]
\centering
\includegraphics[width=0.45\textwidth]{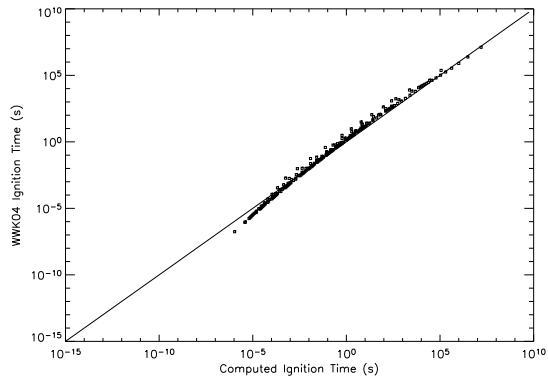}
\caption{ \citet{woosley04} ignition time results vs. calculated
results for constant-pressure ignition delay times for a mixture
half-carbon half-oxygen by mass ($\cfrac = 0.5$), over a
truncated range ($1 \le \rho_8 \le 50$), ($0.5 \le T_9 \le 1.5$). }
\label{fig:woosleyfit}
\end{figure}

The fits are good to within a factor of five between $10^{-9}$~sec and
$1$~sec, and to within a factor of 10 between $10^{-14}$~sec and
$100$~sec.  This can be compared to other analytic expressions, for
instance the constant-pressure formula from \citet{woosley04},
\begin{equation}
{\tau_i} = 15~{\mathrm{sec}} \left ( \frac{7}{T_8} \right )^{22} \left ( \frac{2}{\rho_9} \right )^{3.3}
\end{equation}
which, as shown in Fig.~\ref{fig:woosleyfit}, is an excellent
approximation over a somewhat more narrow range of conditions.

\begin{figure}[htb]
\centering
\plottwo{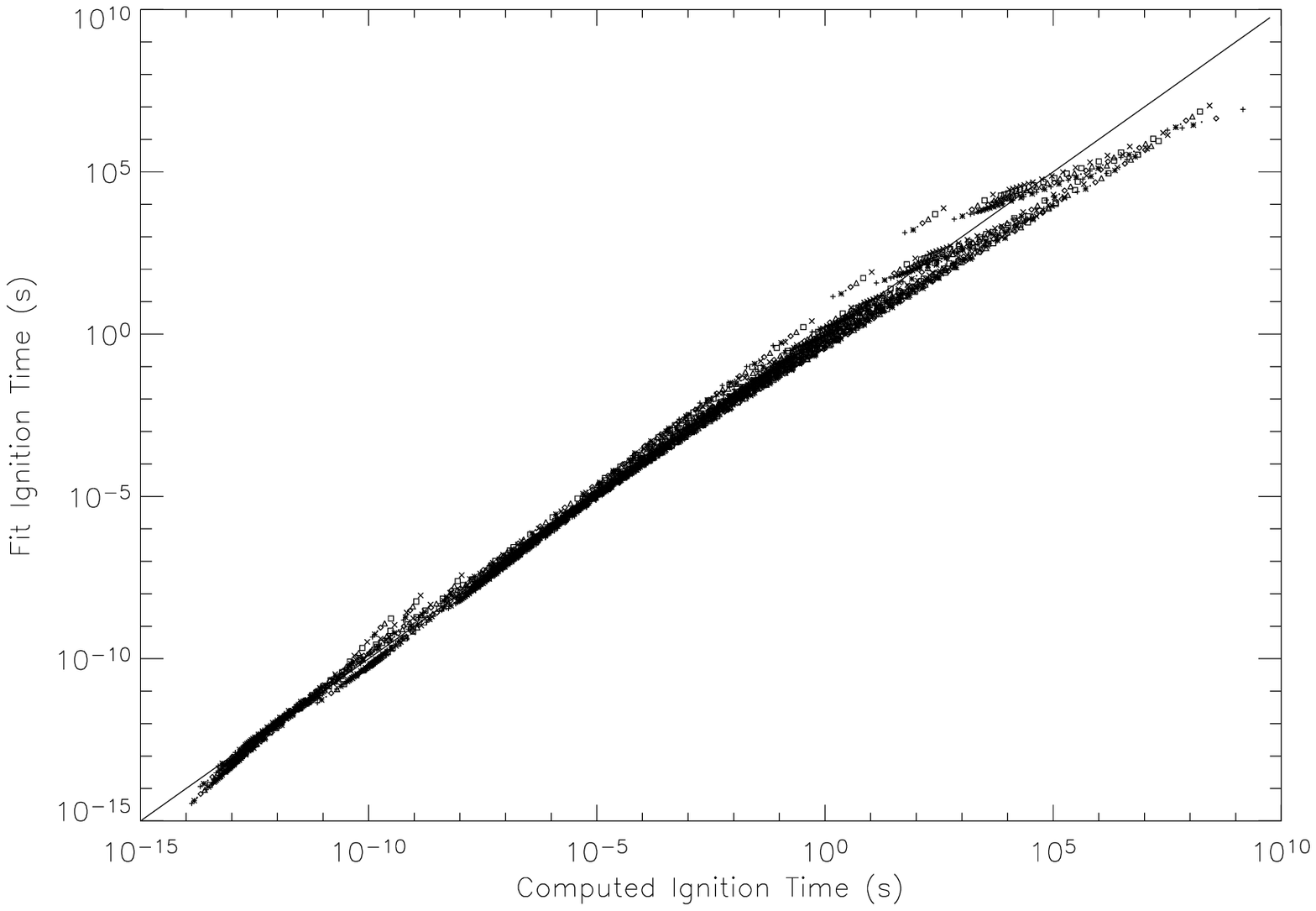}{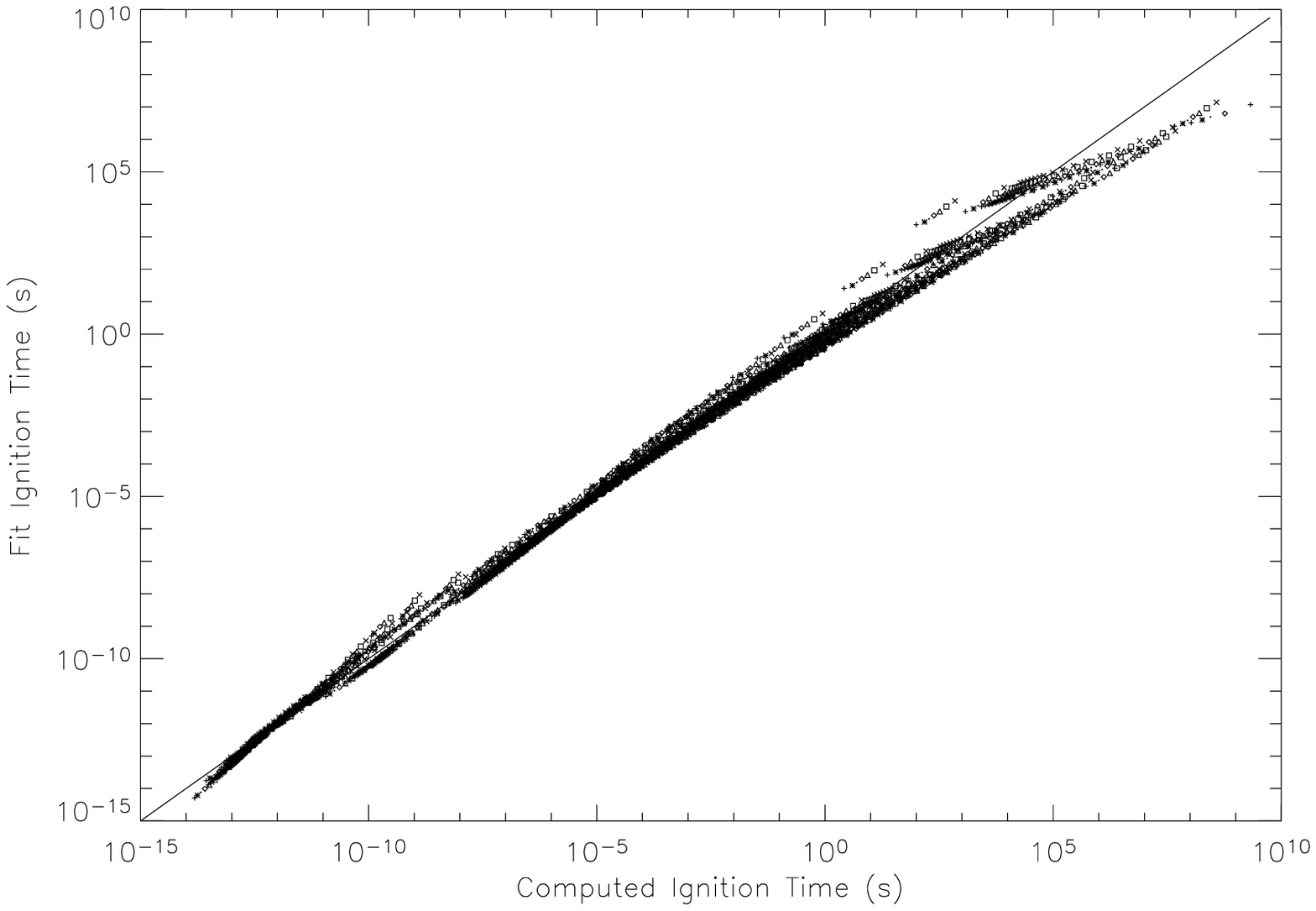}
\caption{ Fit results vs. calculated results for constant-pressure (left)
and constant-volume (right) ignition delay times, for the full range of densities 
($0.1 \le \rho_8 \le 50$), temperatures ($0.5 \le T_9 \le 7$), and 
carbon mass abundances ($0.4 \le \cfrac \le 1.0$) considered.}
\label{fig:fit}
\end{figure}

Which of the two cases (constant volume or constant pressure) are
appropriate will depend on comparing the ignition time with the relevant
hydrodynamical timescale --- in particular, the sound-crossing time of
the region undergoing ignition.   If the region is small enough that many
sound crossing times occur during the ignition, then the constant-pressure
value is appropriate; if ignition occurs in much less than a crossing
time, the constant-volume value is appropriate; intermediate timescales
will result in intermediate values.   In the case of turbulent ignition
of a flame, presumably it is the constant-pressure time which will be
most relevant.   In any case, due to the degeneracy of the material in
the density and temperature ranges considered here, the computed ignition
times (or the fits) rarely differ between the two cases more than 50\%.

\subsection{Effect of Metallicity}

Most of the initial metallicity of main-sequence star comes from the
CNO and \element{Fe}{56} nuclei inherited from its ambient interstellar
medium at birth.  The slowest step in the hydrogen-burning CNO cycle
is proton capture onto \element{N}{14}. This results in all the CNO
catalysts piling up into \element{N}{14} when hydrogen burning on the main
sequence is completed. During helium burning, all of the \element{N}{14}
is converted into \element{Ne}{22}.

As a proxy for investigating the effects of metallicity in our
$\alpha$-chain based reaction network, we consider the ignition of a
$\cfrac = 0.5$ constant-pressure ignition while increasing the fraction
of \element{Ne}{20} (and thus decreasing the abundance of oxygen).
We'll verify this surrogate by using \element{Ne}{22} in larger networks.

\begin{figure}[htb]
\centering
\plottwo{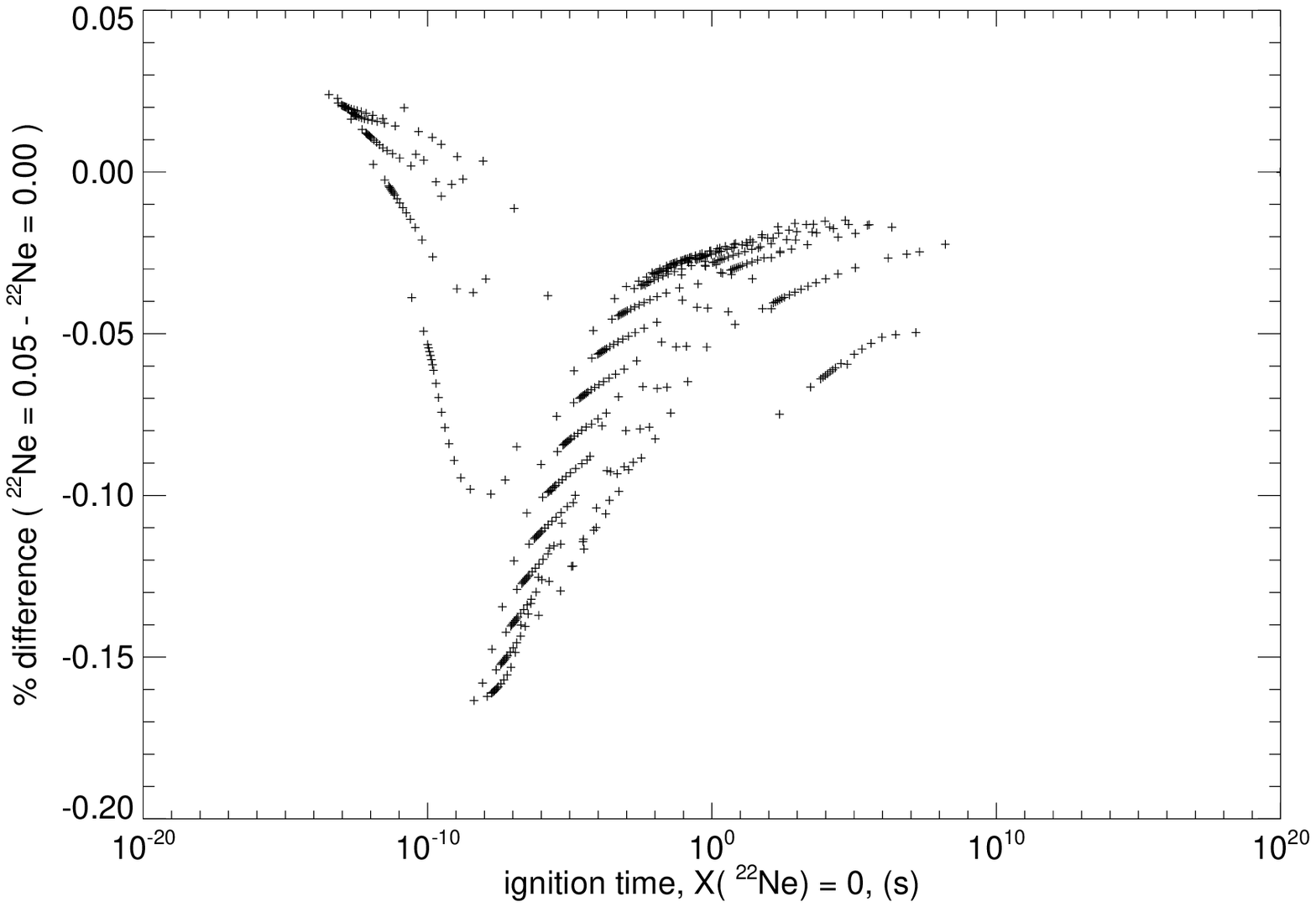}{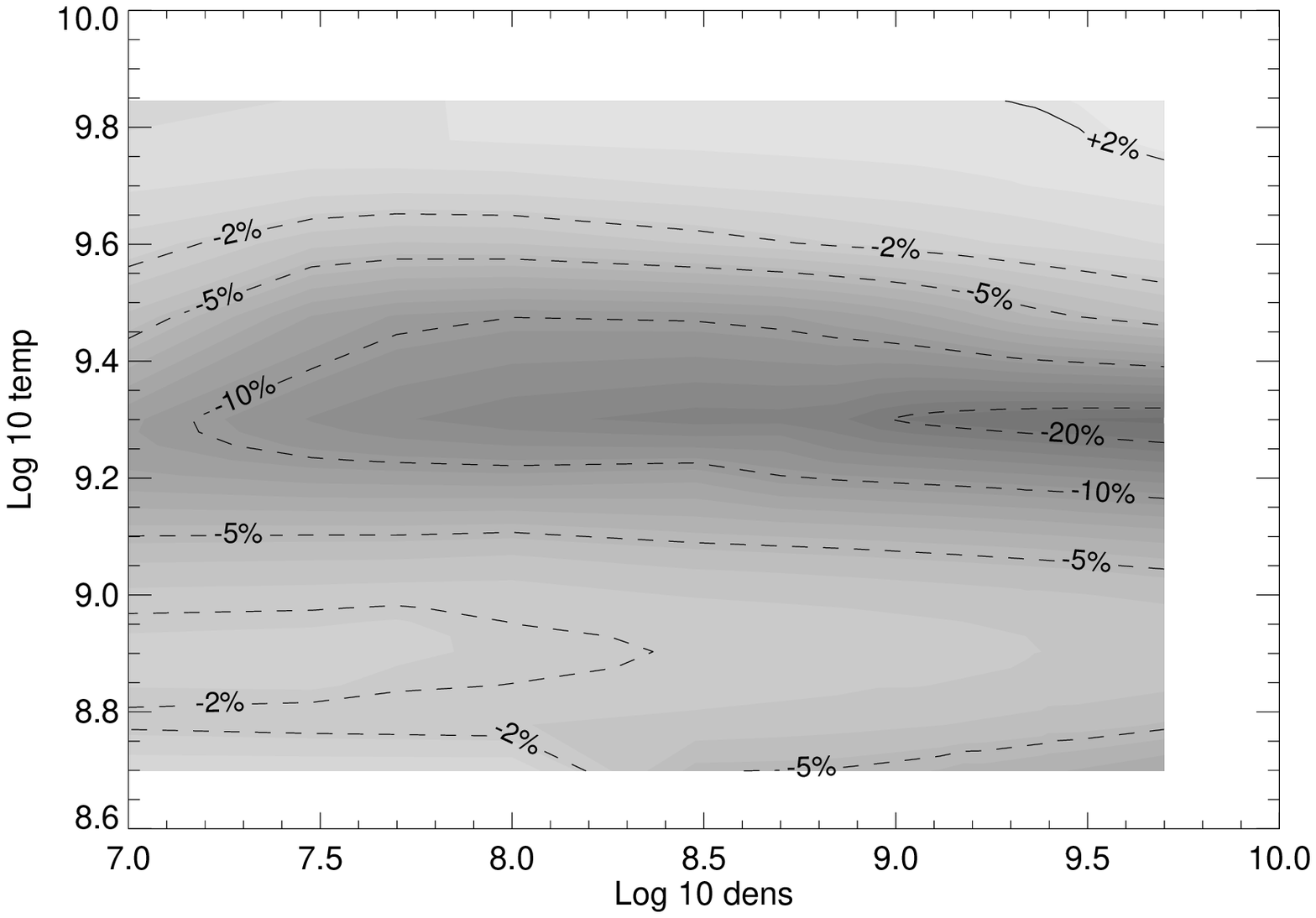}
\setlength{\columnwidth}{\curcolwidth}
\plottwo{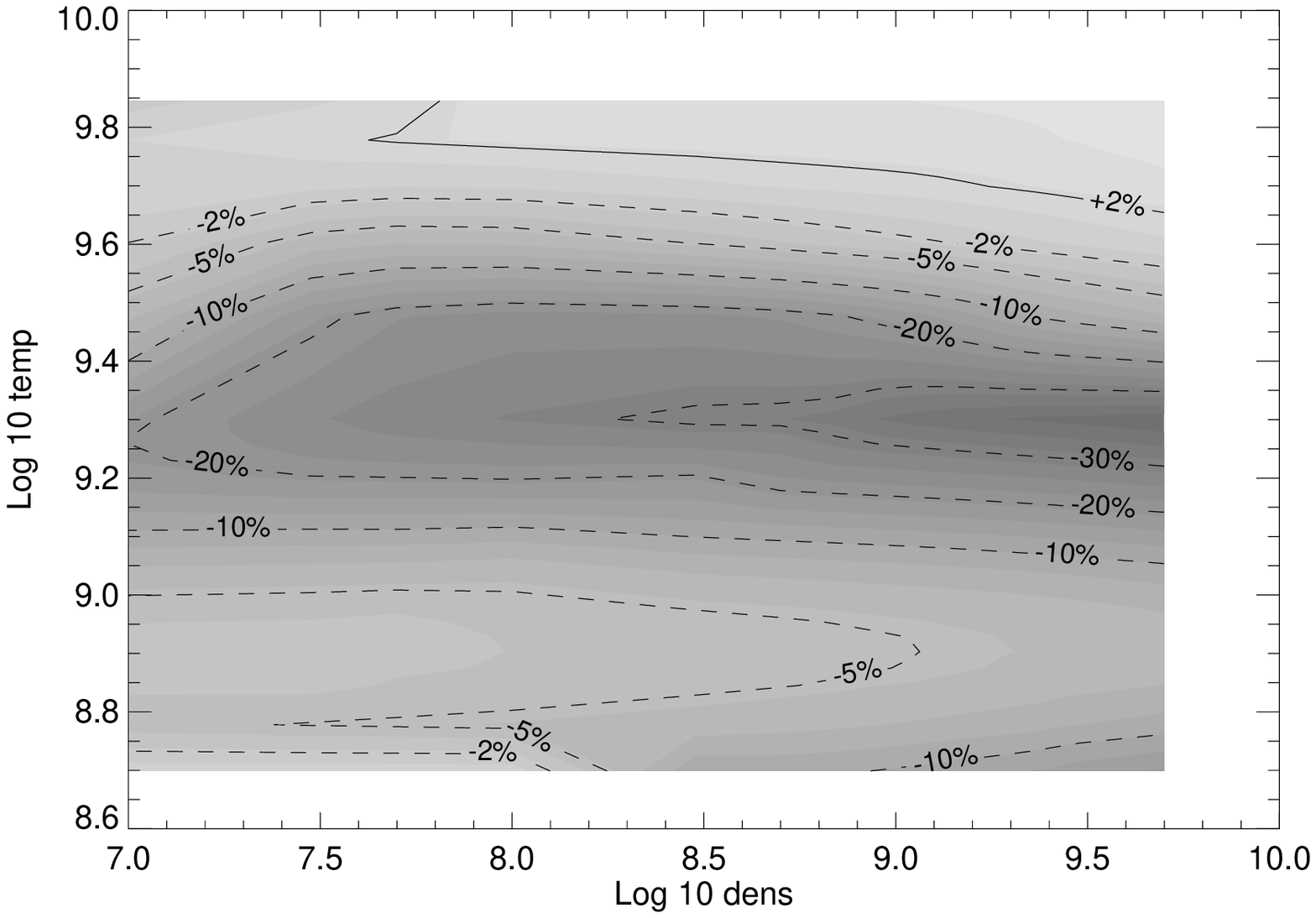}{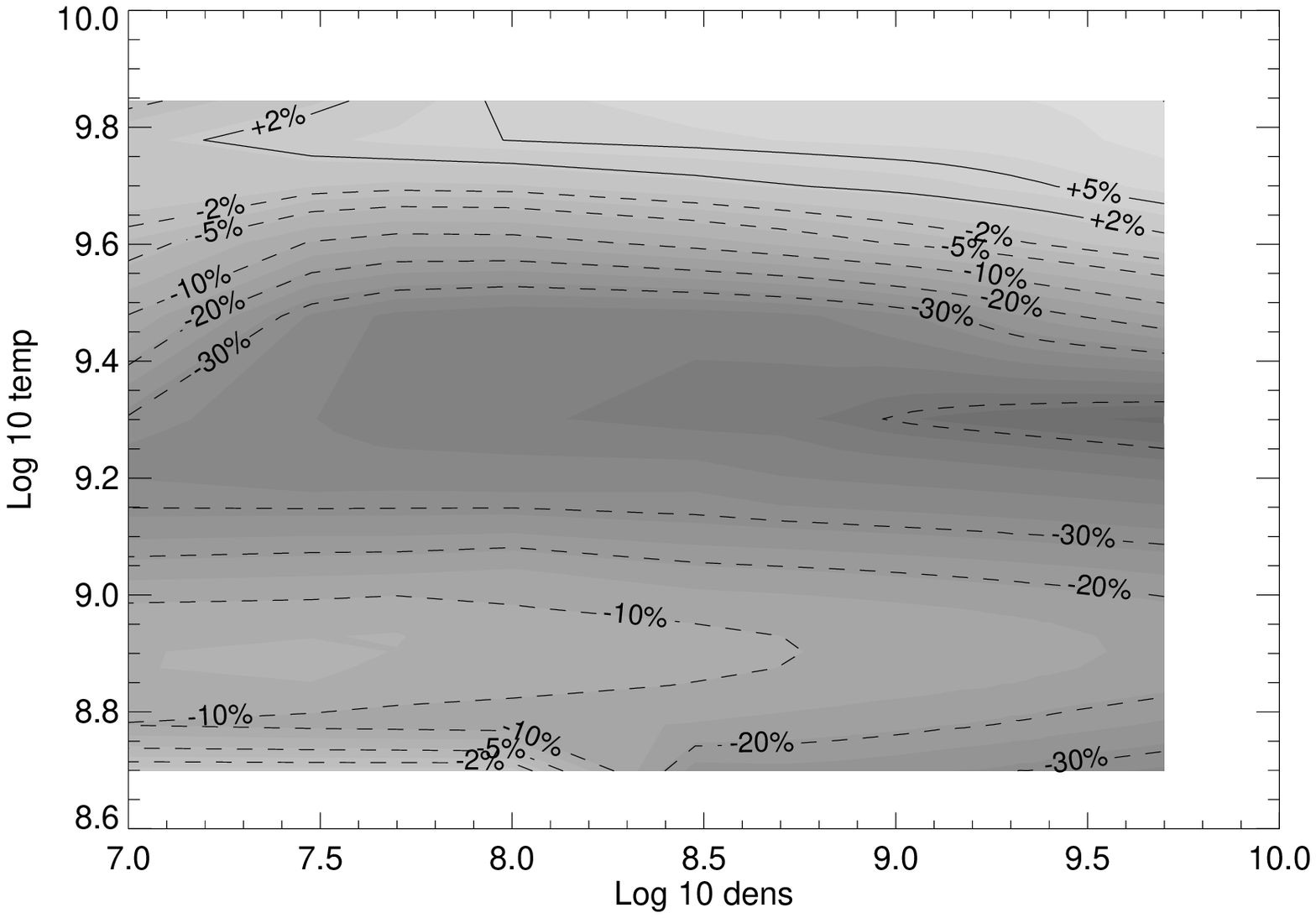}
\caption{The difference in ignition time for a constant-pressure ignition
of $\cfrac =0.5$, $\ofrac=0.5$ when some of the Oxygen is replaced
by Neon-20.   On the top left is shown the fractional difference in
ignition time with the addition of $\netwentyfrac = 0.05$ as a function
of the ignition time with $\netwentyfrac = 0.0$.   On the top right,
bottom left, and bottom right are contour plots in $\rho-T$ space of
the percent difference in ignition time with $\netwentyfrac = 0.05,
0.1, 0.2$, respectively.   The ignition time for the base case is shown
in Fig.~\ref{fig:cp05}.}
\label{fig:nediff}
\end{figure}

The effects of increasing \netwentyfrac{} from 0 to 0.05, and then further
to 0.1 and 0.2, is shown in Fig.~\ref{fig:nediff}.  Addition of even
fairly modest amounts of neon can significantly (20--30\%) reduce the
ignition times for much of the thermodynamic conditions evaluated here.
The reduced ignition times are found to result from a larger energy
release from burning over the entire integration range.

To understand the increase in the energy deposited by burning,
consider the $\alpha$-chain reactions which dominate burning in this
regime and are modeled by the `aprox13' network as:

\begin{eqnarray}
\element{C}{12}  + \element{C}{12}  & \rightarrow & \element{Ne}{20} + \alpha + 13.933 \, \MeV \nonumber \\
\element{C}{12}  + \alpha  & \rightarrow &  \element{O}{16}      + 7.162 \, \MeV \nonumber \\
\element{O}{16}  + \alpha  & \rightarrow &  \element{Ne}{20}     + 4.734 \, \MeV \\
\element{Ne}{20} + \alpha  & \rightarrow &  \element{Mg}{24}     + 9.312 \, \MeV  \nonumber \\
\element{Mg}{24} + \alpha  & \rightarrow &  \element{Si}{28}      + 9.984 \, \MeV. \nonumber
\end{eqnarray}

Given an initial mixture of \element{C}{12}, \element{O}{16}, and
\element{Ne}{20}, it is the carbon burning reaction
$\element{C}{12}+\element{C}{12}$ which happens first.  The resulting
$\alpha$-particle can capture onto carbon or any heavier element in
the $\alpha$ chain. Unless that chain is already populated, then the
(very exothermic) neon capture and all flows to still heavier nuclei
are choked off until enough \element{Ne}{20} and \element{Mg}{24} are
generated through burning.  Adding even quite modest amounts of neon
to the initial mixture allows more $\alpha$-chain reactions to
promptly occur during carbon burning.  

\begin{figure}[htb]
\centering
\plottwo{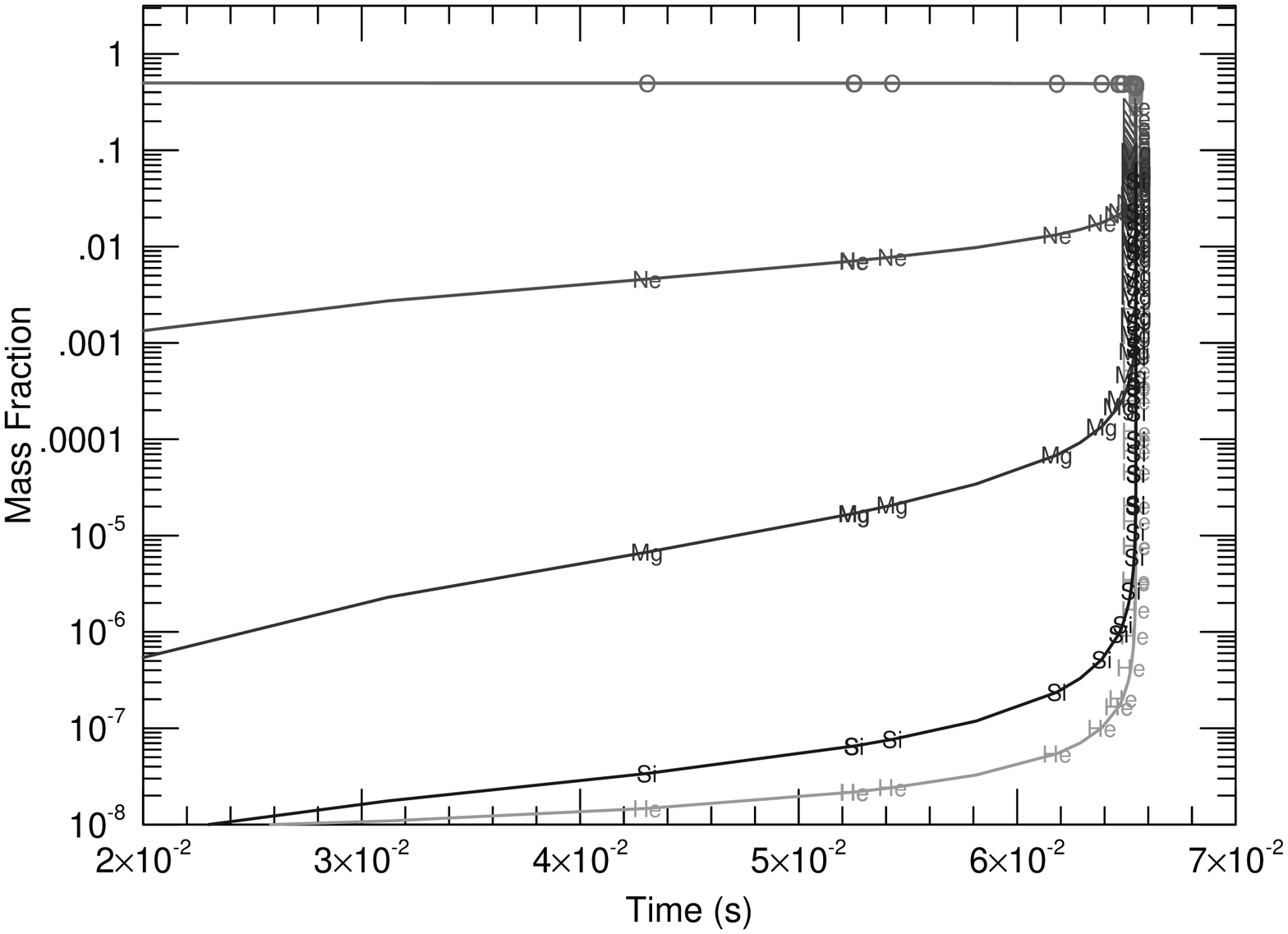}{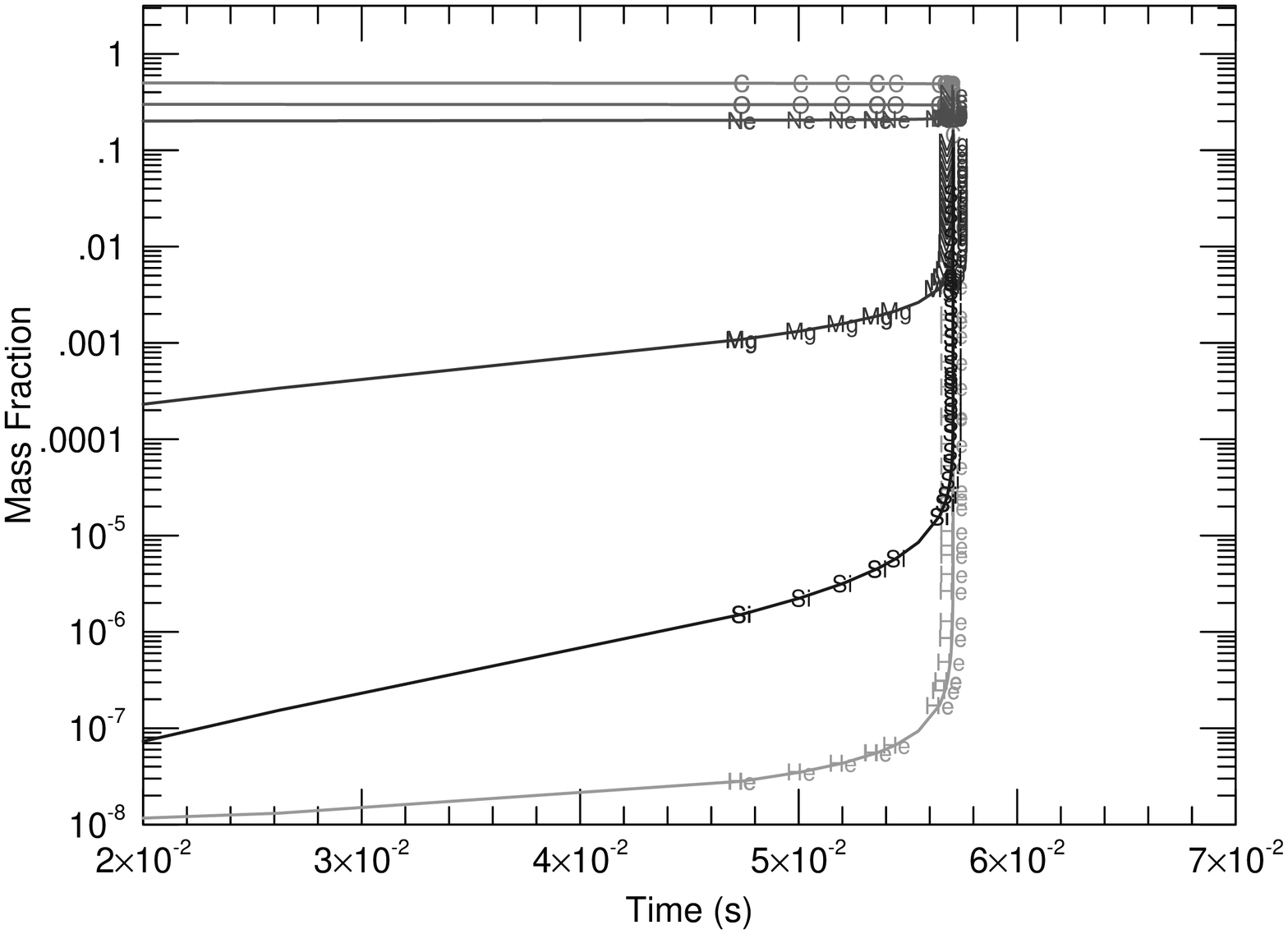}
\caption{Mass fraction evolution in constant-density burning, with
an initial state of $\rho_8 = 10$, $T_9 = 1$, $\cfrac = 0.5$, with
$\netwentyfrac = 0.0$ (left) and $\netwentyfrac = 0.1$ (right).  The
burning here was calculated with a 513-isotope network. 
Because of the removal of the $\alpha$-chain bottleneck at neon on the right,
burning proceeds faster and the generation of higher intermediate-mass
elements is raised by orders of magnitude at early times.}
\label{fig:abundanceevolution}
\end{figure}

The plots in Fig.~\ref{fig:abundanceevolution} show the abundance
evolution of $\alpha$-chain elements during constant-density burning at
$\rho_8 = 10$, $T_9 = 1$, $\cfrac = 0.5$, and an initial neon abundance
of zero and 0.05.  The inclusion of 5\% neon by mass greatly speeds
the production of heavier intermediate-mass elements, and thus the
exothermicity of the burning, reducing the ignition time.

To confirm that this effect is not artificially enhanced by using
an $\alpha$-chain reaction network, and to quantitatively verify the
ignition times produced by the aprox13 network used here, we compared
the ignition times at two $(\rho,T)$ points and varying neon abundance
with those produced by reaction networks containing 513 and 3304
isotopes.  The results are shown in Table~\ref{tab:networkcompare1} 
We see that not only are the times computed
with the smaller network quantitatively in good agreement with the larger
networks (within 10\% for the lower-temperature case, and within 25\%
for the higher case), but the trends are similar.  The same trend is
apparent when \element{Ne}{22}, unavailable in the $\alpha$-chain network,
is used instead of \element{Ne}{20}.  The trend is in fact stronger with
\element{Ne}{22} because of additional flow paths that become available.

\begin{deluxetable}{ccccccccc}
\tablecaption{Constant-volume ignition times (sec) adding Neon with different reaction networks}
\tablecolumns{9}
\tablewidth{0pt}
\tabletypesize{\footnotesize}
\tablehead{ & & & & \multicolumn{3}{c}{$\rho_8 = 10, T_9 = 1$} & \multicolumn{2}{c}{$\rho_8 = 1, T_9 = 3$} \\
       \colhead{\cfrac} &  \colhead{\ofrac} & \colhead{\netwentyfrac} & \colhead{\netwentytwofrac} &  
       \colhead{13 isotopes\tablenotemark{a}} & \colhead{513 isotopes} & \colhead{3304 isotopes} & 
       \colhead{13 isotopes\tablenotemark{b}} & \colhead{513 isotopes}}
\startdata
$0.5$ &   $0.5$   &   $0.0$  &  $0.0$  & $5.96 \times 10^{-2}$ & $6.42 \times 10^{-2}$  & $6.42 \times 10^{-2}$ &
                                         $2.16 \times 10^{-8}$ & $1.74 \times 10^{-8}$  \\
$0.5$ &   $0.45$  &   $0.05$ &  $0.0$  & $5.77 \times 10^{-2}$ & $6.49 \times 10^{-2}$  & &
                                         $1.94 \times 10^{-8}$ & $1.58 \times 10^{-8}$  \\
$0.5$ &   $0.4$   &   $0.1$  &  $0.0$  & $5.58 \times 10^{-2}$ & $6.38 \times 10^{-2}$  & $6.38 \times 10^{-2}$ &
                                         $1.79 \times 10^{-8}$ & $1.49 \times 10^{-8}$  \\ 
$0.5$ &   $0.3$   &   $0.2$  &  $0.0$  & $5.18 \times 10^{-2}$ & $6.09 \times 10^{-2}$  & &
                                         $1.60 \times 10^{-8}$ & $1.37 \times 10^{-8}$  \\
$0.5$ &   $0.45$  &   $0.0$  &  $0.05$ &           & $5.56 \times 10^{-2}$  & & &
                                        $9.61 \times 10^{-9}$  \\
$0.5$ &   $0.4$   &   $0.0$  &  $0.1$  &           & $5.54 \times 10^{-2}$  & $5.54 \times 10^{-2}$ & &
                                        $9.04 \times 10^{-9}$  \\
$0.5$ &   $0.3$   &   $0.0$  &  $0.2$  &           & $5.64 \times 10^{-2}$  &  & &
                                        $8.92 \times 10^{-9}$  \\
\enddata
\tablenotetext{a}{Ignition time in seconds.  The `aprox13'
network used primarily in this work gives ignition times within 10\%
of more complete networks for this $(\rho,T)$, and with similar trends.}
\tablenotetext{b}{Ignition time in seconds.  The `aprox13'
network used primarily in this work gives ignition times within 25\%
of more complete networks for this $(\rho,T)$, and with similar trends.}
\label{tab:networkcompare1}
\end{deluxetable}

\begin{figure}[htb]
\centering
\plottwo{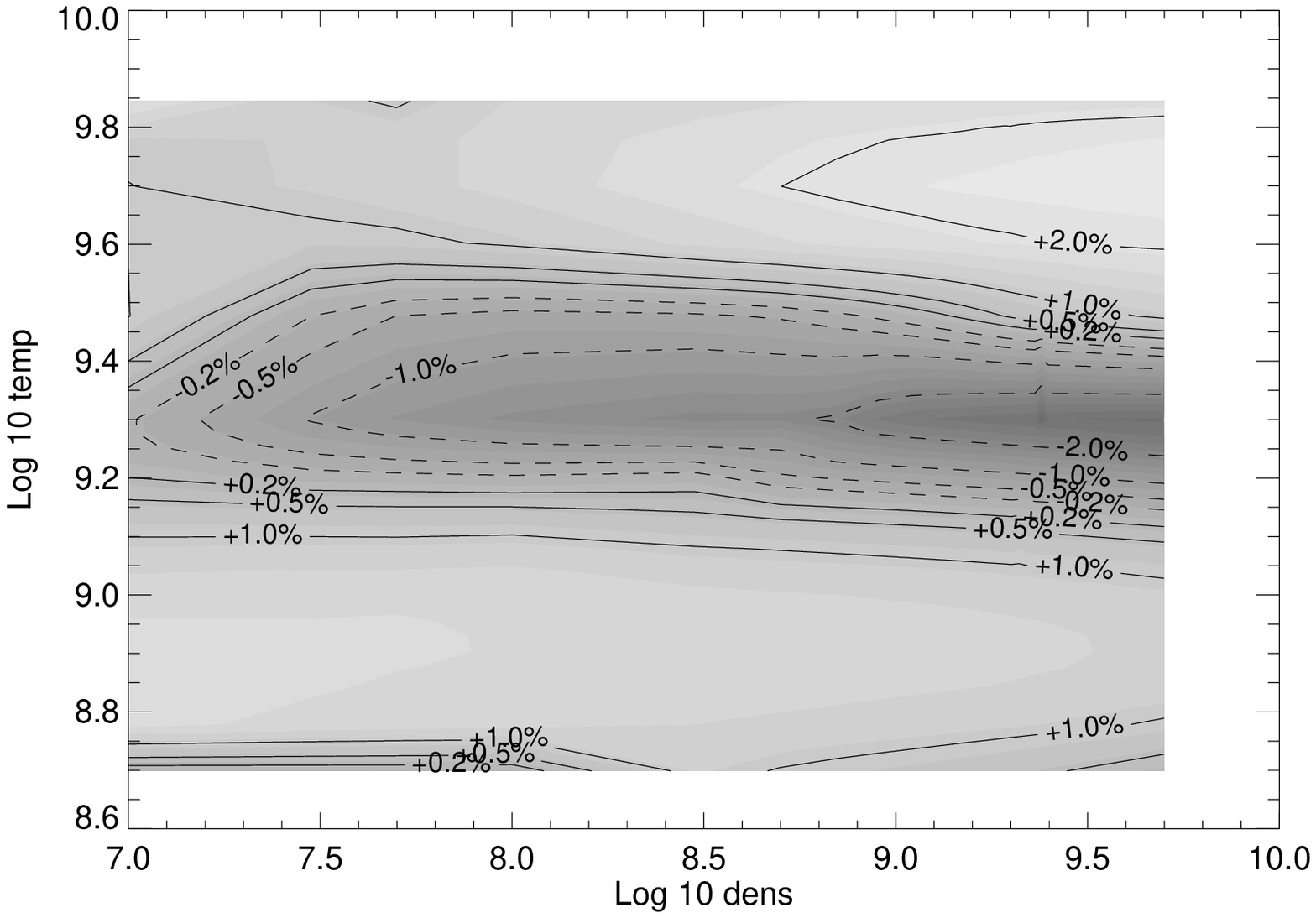}{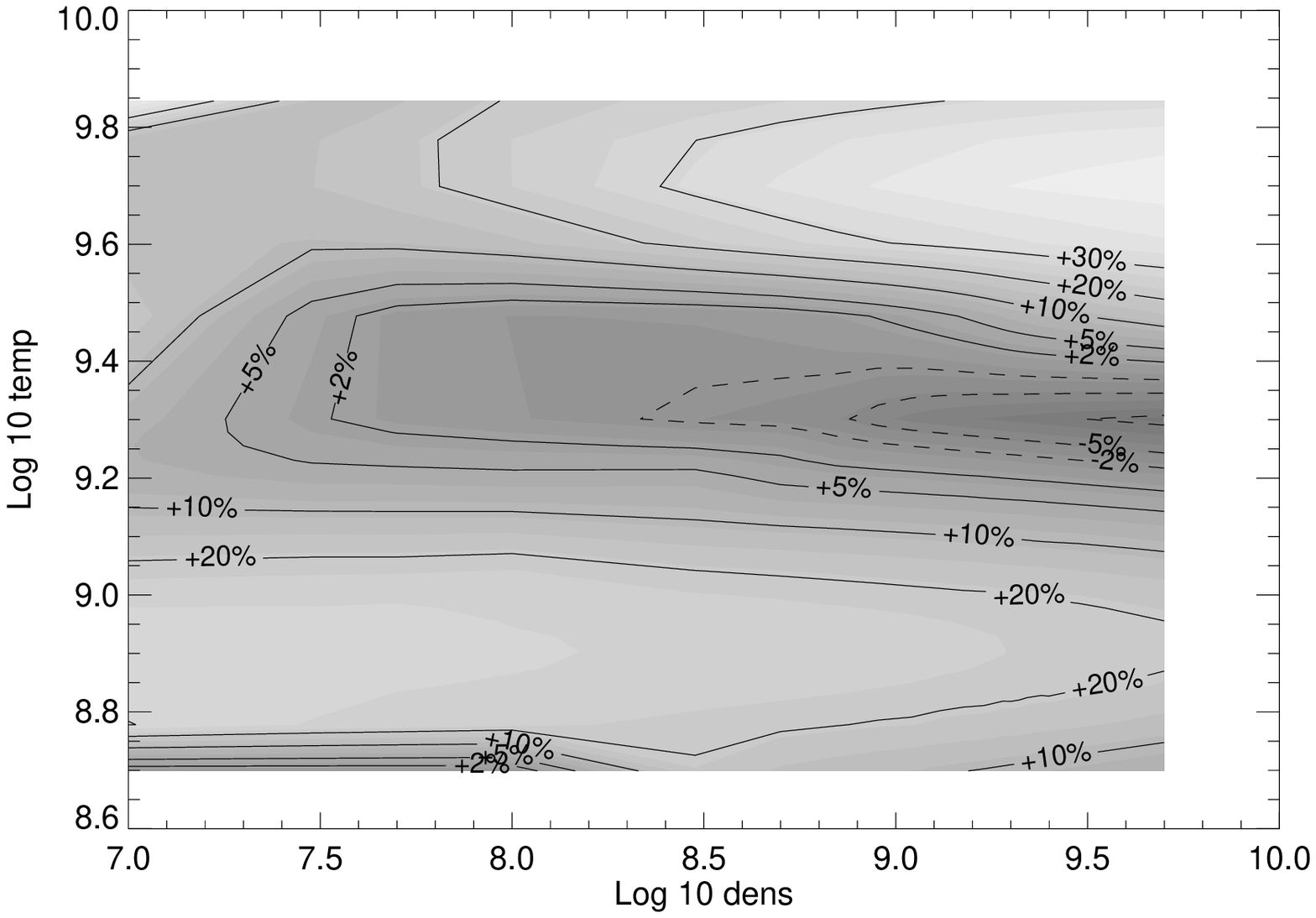}
\caption{As in Fig.~\ref{fig:nediff}, the difference in ignition time for a constant-pressure ignition
of $\cfrac =0.5$, $\ofrac=0.5$ when (left) a mass fraction of 0.005 of each of the carbon and oxygen is replaced
by Neon-20 and (right) when 0.05 of each is replaced by Neon.  Note that in this case, for $\netwentyfrac = 0.1$, with
the exception of a small region in ($\rho$,$T$) the ignition time is increased by the same magnitude that it
is decreased in the case when all of the neon comes from oxygen, \eg{} the $\cfrac=0.5$,$\ofrac=0.4$,$\netwentyfrac=0.1$
case of Fig~\ref{fig:nediff} in the bottom right panel.}
\label{fig:nediff2}
\end{figure}

While the above has demonstrated the active role neon plays in the
ignition process, its effect is much smaller than that of carbon, which
is the primary source of fuel in this burning process.   As a result, in
the more realistic case when an increase in neon comes at the expense of
both carbon and oxygen, this effect is reduced (for small neon fractions)
or completely reversed (for moderate neon fractions).  This is shown, for
instance, in Fig.~\ref{fig:nediff2}.   In this case, increased metallicity
of the progenitor system has the opposite effect; it makes the ignition
time significantly longer for hotspots in the resulting white dwarf.

\section{APPLICATION: IGNITION OF A DETONATION}

\subsection{Detonation Structure}

Detonation waves are a supersonic mode of propagating combustion.
A shock wave heats up material, which then ignites, releasing energy
which further powers the shock.  \citep[See, for instance, textbooks such
as][]{glassman96,williams}.  There are, broadly, four
states in a detonation: the unshocked material; the shocked material
immediately behind the shock; an induction zone, where the heated material
slowly begins burning and then the reaction zone, where the bulk of the
exothermic burning takes place.

Unsupported, self-sustaining detonations can be of the Chapman-Jouget
(CJ) type, where at the end of the reaction zone the flow becomes
sonic, or of the pathological type, where the sonic point occurs within
reaction zone, decoupling the flow downstream of the sonic point from
the shock.  Pathological detonations can occur in material where there
are endothermic reactions or other dissipative or cooling effects, and
may have speeds slightly higher (typically by a few percent) than the
CJ speed.  Detonations within highly degenerate white-dwarf material
($\rho_8 > 0.2$) are of the pathological type \citep{khokhlov89}, largely
because some regions of the flow behind the detonation can have
significant amounts of endothermic reactions, violating the assumptions
of the CJ structure.  Because in the case of a pathological detonation
some fraction of the reactions powering the detonation are decoupled from
the shock, calculating the speed of a pathological detonation requires
detailed integration of the detonation structure, rather than simply
using jump conditions.  For either kind of detonation, one can estimate
where most burning occurs with shock speed $D$ (which, even in non-CJ
case, can be estimated with CJ speed) and $\tau_i$; $l_i = D \tau_i$
is the position behind the shock at which the induction zone ends and
rapid burning takes place.

\begin{figure}[htb]
\centering
\plottwo{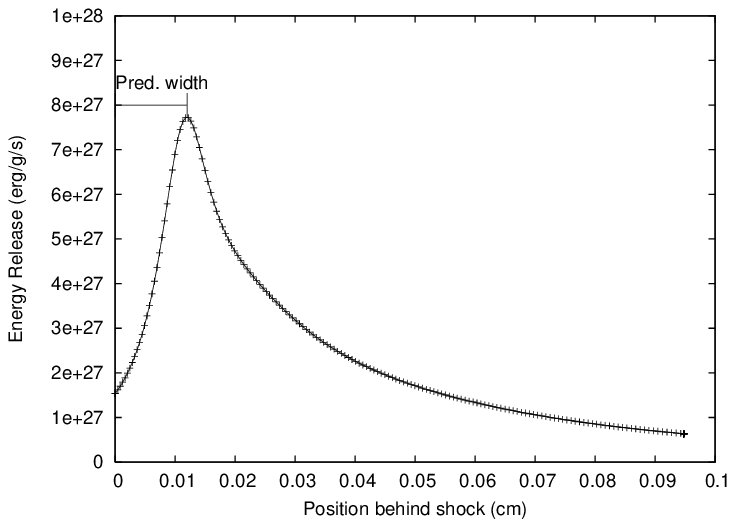}{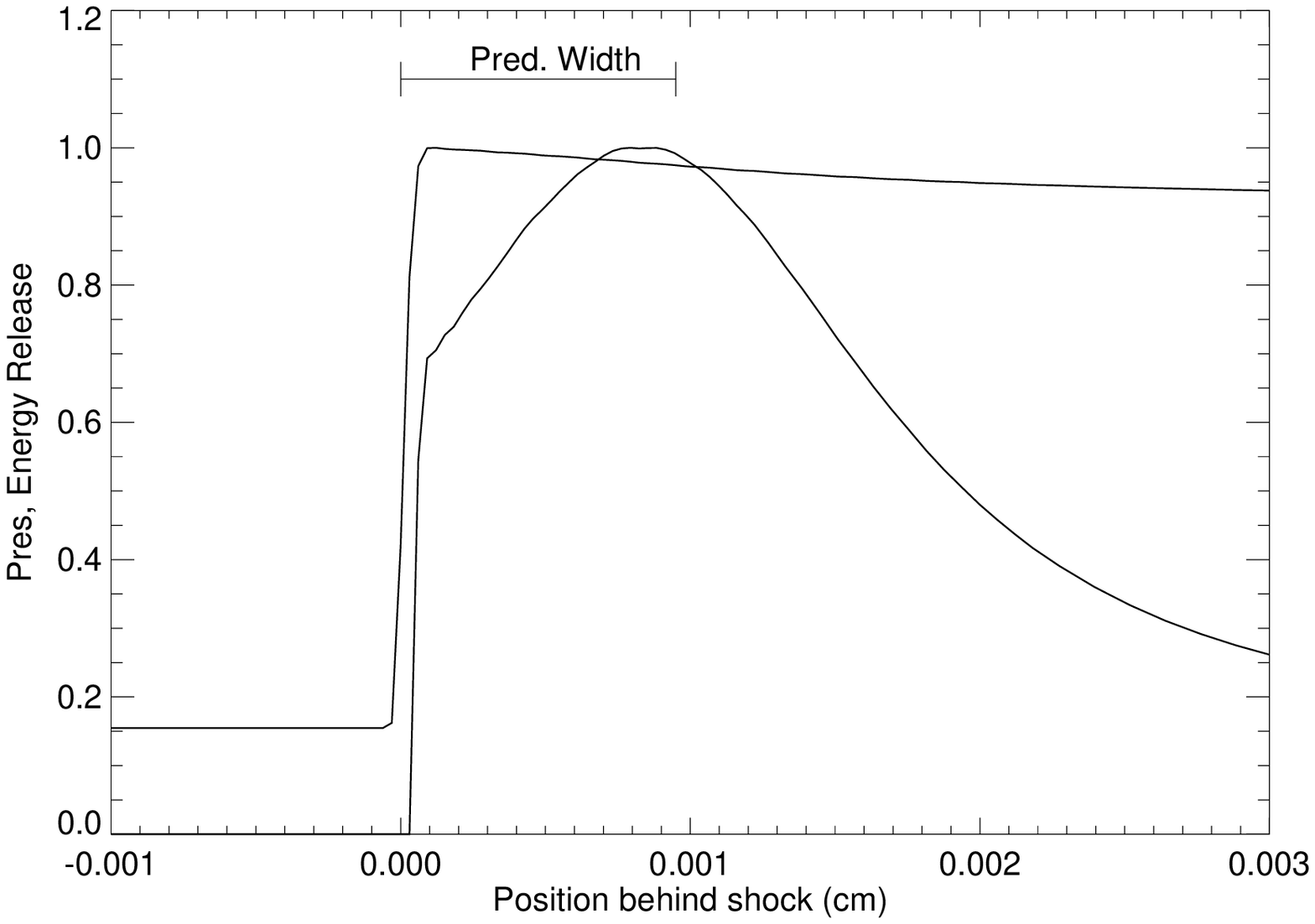}
\caption{Two examples of estimating the detonation thickness, $l_i = v_s \tau_i$
in a detonation.   On the left, plotted energy release rate from nuclear
reactions behind the shock of a leftward-traveling ZND detonation
into a pure-carbon quiescent medium of $\rho_8 = 1$, $T_9 = 0.05$.
The shocked state is $\rho_8 = 2.97$, $T_9 = 4.2$, and the incoming fuel
velocity behind the shock is $4.0 \times 10^8 \cms$.  For the shocked
material, the predicted ignition time is $\approx 3 \times 10^{-11}
\s$.  Even in this case, where the shocked temperature is so high that
significant burning occurs immediately, and the `square wave' detonation
structure does not apply, the predicted $l_{i} = 1.2 \times 10^{-2}
\cm$ correctly matches the peak of the reaction zone.   On the right,
pressure (top) and energy release rate (bottom), plotted relative to their maximum values,
behind a leftward-traveling slightly overdriven detonation into the
same material as in the previous figure, calculated by the hydrodynamics
code \FLASH{} \citep{flashcode,flashvalid}.  The line above the plotted
quantities shows the predicted $l_{i}$ calculated with the observed
values in the shocked state. }
\label{fig:detstructure}
\end{figure}

In the case of a detonation into a very low-density, cold material, the
material immediately behind the shock will still not burn significantly
until a length of time equal to the ignition time passes, resulting in a
`square wave' detonation.  For conditions relevant to near the core
of a white dwarf, however, the post-shock fluid will typically have
temperatures on order $T_9 \approx 5$ -- that is, temperatures which
are already near the maximum temperature which will be obtained by burning.
Even in these cases, this estimate of $l_i$ provides a good measure
of the thickness of the detonation structure behind the shock, as is
shown in Fig~\ref{fig:detstructure} where it measures the position of maximum burning.

\subsection{Ignition of a Spherical Detonation}

One mechanism for ignition of a detonation by a hotspot is an initial
rapid input of energy which leads to a Sedov blast wave; the material
shocked by the outgoing spherical blast ignites, and as the outgoing
wave slows, a steady outgoing detonation results.  A steady detonation
cannot form until the outgoing shock wave speed drops to the detonation
velocity detonation, or else the energy released by reactions behind
the blast wave will not be able to `catch up' to the outgoing shock to
drive it.   On the other hand, if  the shock drops significantly below
the detonation speed before ignition takes place, not enough material
will be burning per unit time to sustain a detonation wave.  We denote
the position of the shock when it reaches the detonation speed from above
as $R_{D}$.  Naively, the condition for successful detonation ignition
would be that $R_{D} \gtrsim l_{i}$, since a detonation structure with
width of order $l_{i}$ must be set up before the shock speed becomes
too slow; a more sophisticated derivation of this criterion can be
found in \citet{zeldovich56}.  However, experimentally this is known to
be far too lenient a condition for terrestrial detonations and $R_{D}$
must be orders of magnitude larger than $l_{i}$ \citep{desbordes86}.
This has also been empirically seen in the context of astrophysical
detonations \citeeg{niemeyerwoosley97}.

This has been explained by, for instance, \citet{heclavin}.  Curvature has
a significant nonlinear effect on the structure of a detonation --- even
more so than on the structure of a flame \citep[\eg{},][]{flamecurve}
because the curvature directly effects the burning region rather than 
merely the preconditioning (diffusion) region.   \citet{heclavin},
looking at a pseudo-steady calculation of a detonation with curvature,
find that for a steady detonation to exist requires the curvature to be
extremely small.  The condition found by the authors requires $R_{D}
\approx 44 \gamma^2/(\gamma^2 - 1) \beta_Z l_{i}$, where $\gamma$
is the polytropic index of the ideal gas and $\beta_Z$ represents
the temperature sensitivity of the burning law; for the reactions and
temperatures of interest in astrophysical combustion, $\beta_Z \approx
10-15$ \citep{flamecurve}.  Using $\gamma = 4/3$ to describe the highly
degenerate material near the centre of the white dwarf, this would give
$R_D \approx 1000 - 1500 l_i$.

While in many cases using a polytropic ideal gas equation of state
to describe degenerate white dwarf material can be an excellent
approximation, in combustion phenomenon where the burning rate is
highly temperature dependent it is problematic, making the above
result unsuitable.   Further, the authors assume a CJ detonation, as
opposed to the pathological detonation that occurs at high densities in
a white dwarf.  This makes a straightforward application of the 
results of \citet{heclavin} to our problem of interest difficult.
Given the complexity of the partially degenerate equation of state in
the white dwarf, re-deriving the analytic results for this application
would be difficult.   However, the detailed effect of curvature on detonations
in white dwarfs has already been studied in a different manner, by
\citet{sharpecurved}.   In this approach, the velocity eigenvalue for
the detonation -- which depends sensitively on the detonation structure
-- is calculated using a shooting method to numerically integrate the
structure of a pathological detonation to the sonic point.  A range of
possible detonation velocities is input, and then repeatedly bisected as
the detonation structure with a fixed given curvature term is integrated
assuming the current detonation velocity.   The result is the
velocity-curvature relation for a detonation into a given ambient material,
and as a byproduct, the relationship between the thermodynamic structure
(at least up to the sonic point) and the curvature for the curved detonation.

\begin{figure}[htb]
\centering
\plotone{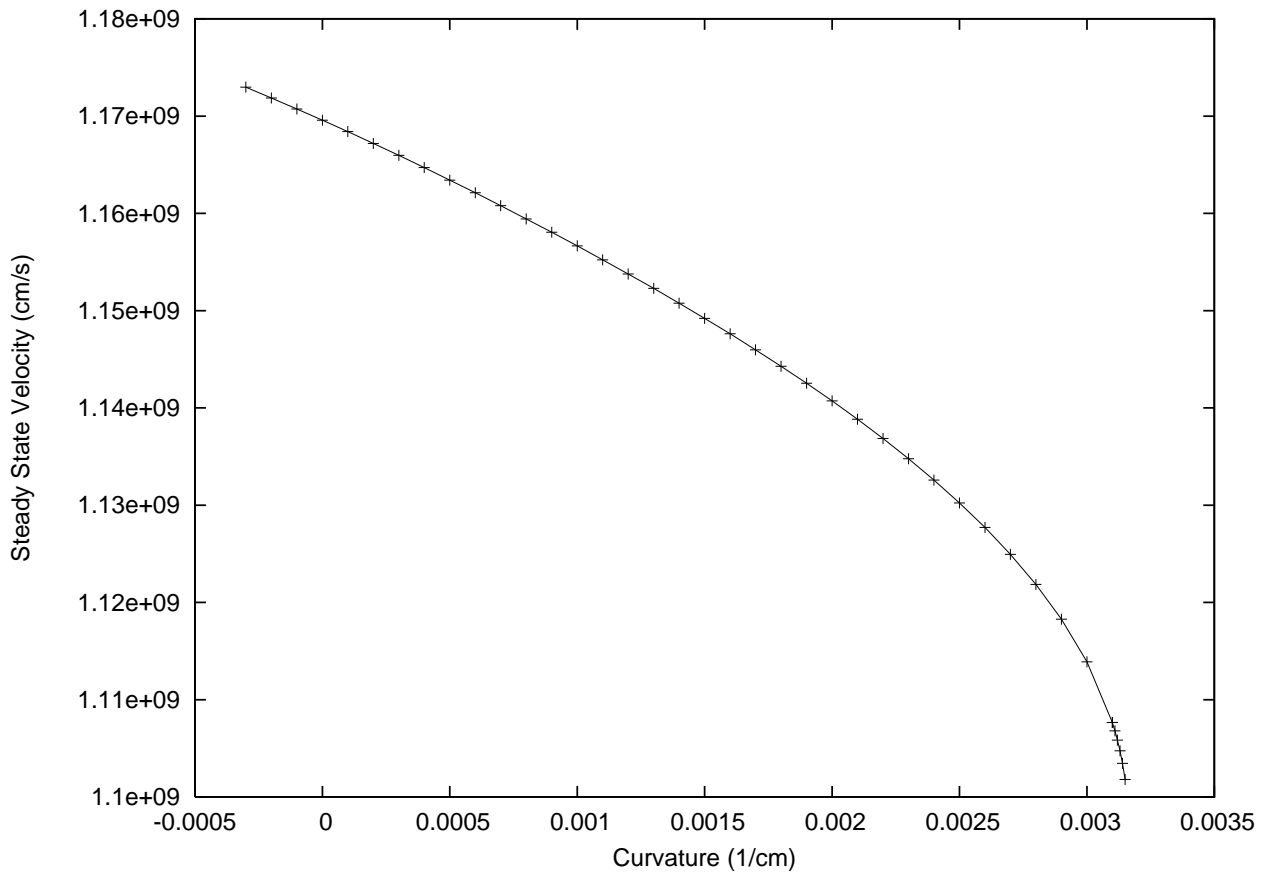}
\caption{Example of detonation speed vs. curvature, for a detonation into
a quiescent medium of $\rho_8 = 1$, $T_9 = 0.05$, $\cfrac = 0.5$, $\ofrac = 0.5$.}
\label{fig:detcurvevelexample}
\end{figure}

We follow the method of \citet{sharpecurved}, also described in
Appendix~\ref{sec:steadystatedet}, and measure the maximum sustainable
spherical curvature of a detonation using the same equation
of state and burning network as used in the ignition time study.
An example of the detonation-speed versus curvature relation is given
in Fig.~\ref{fig:detcurvevelexample}.   Beyond some maximum curvature
$\kmax$, no steady-state unsupported detonation can exist; thus in the
case of a spherical detonation, for a steady detonation to successfully
ignite, $R_D > \kmax^{-1}$.

\begin{figure}[htb]
\centering
\plottwo{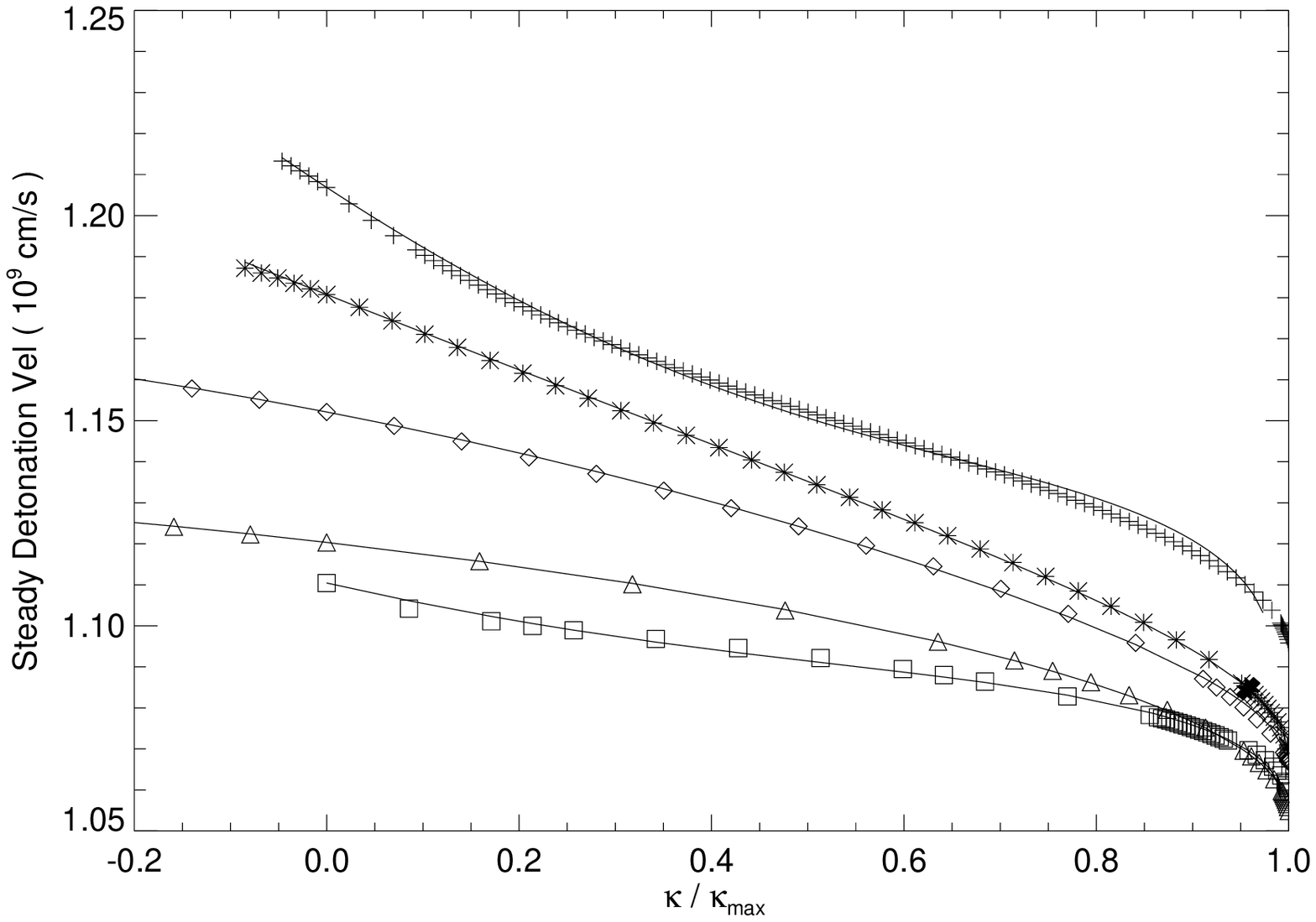}{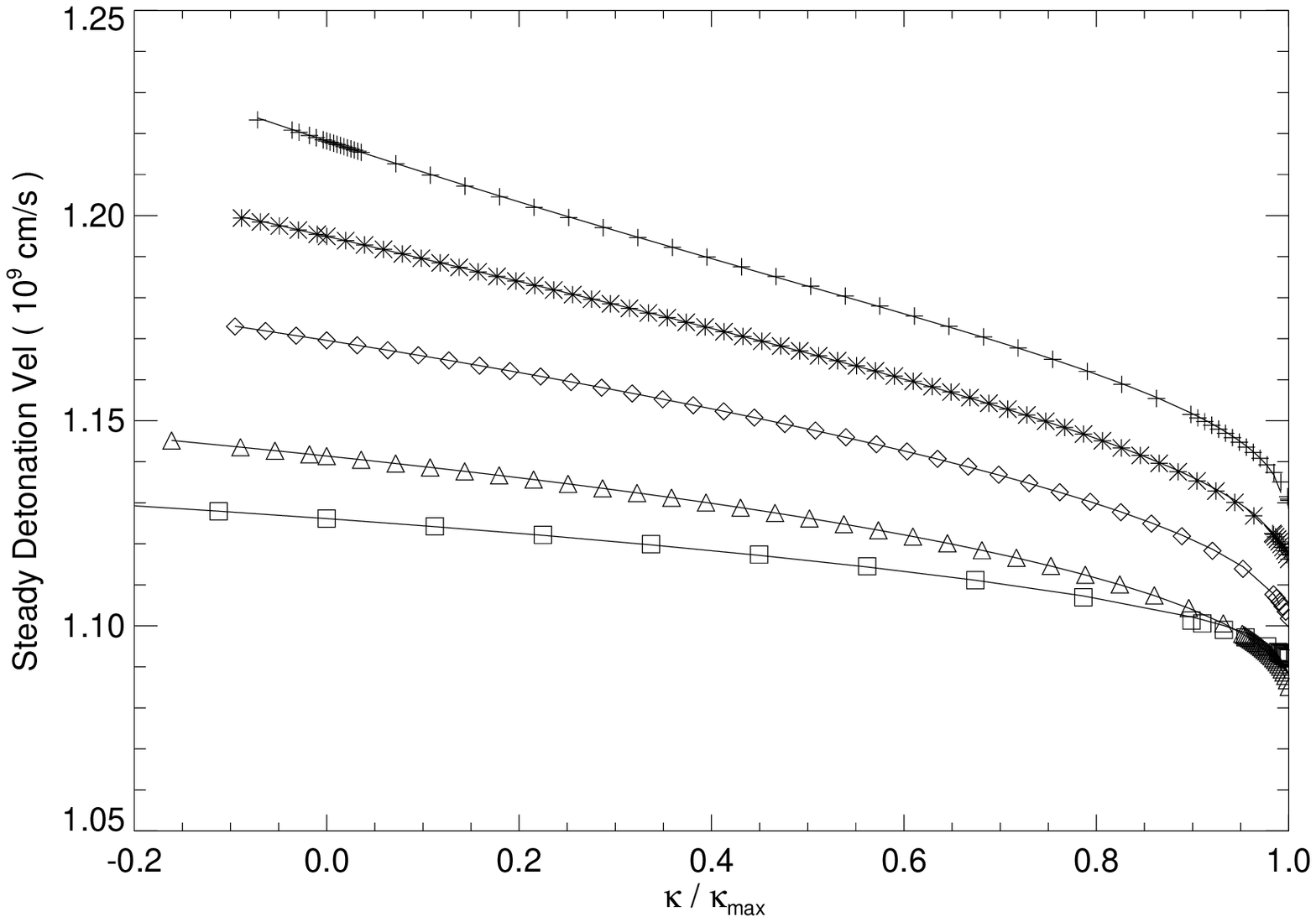}
\setlength{\columnwidth}{\curcolwidth}
\plottwo{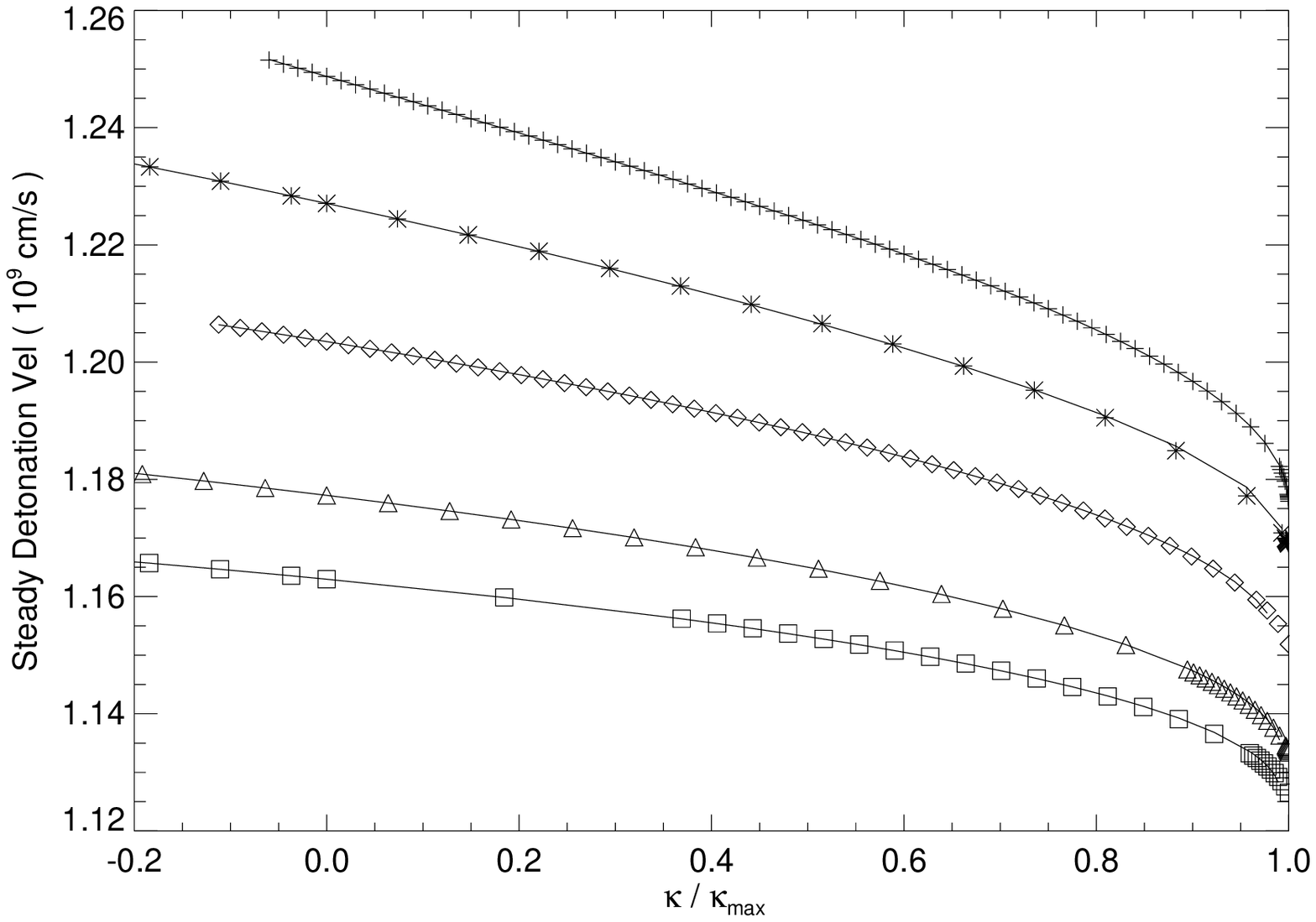}{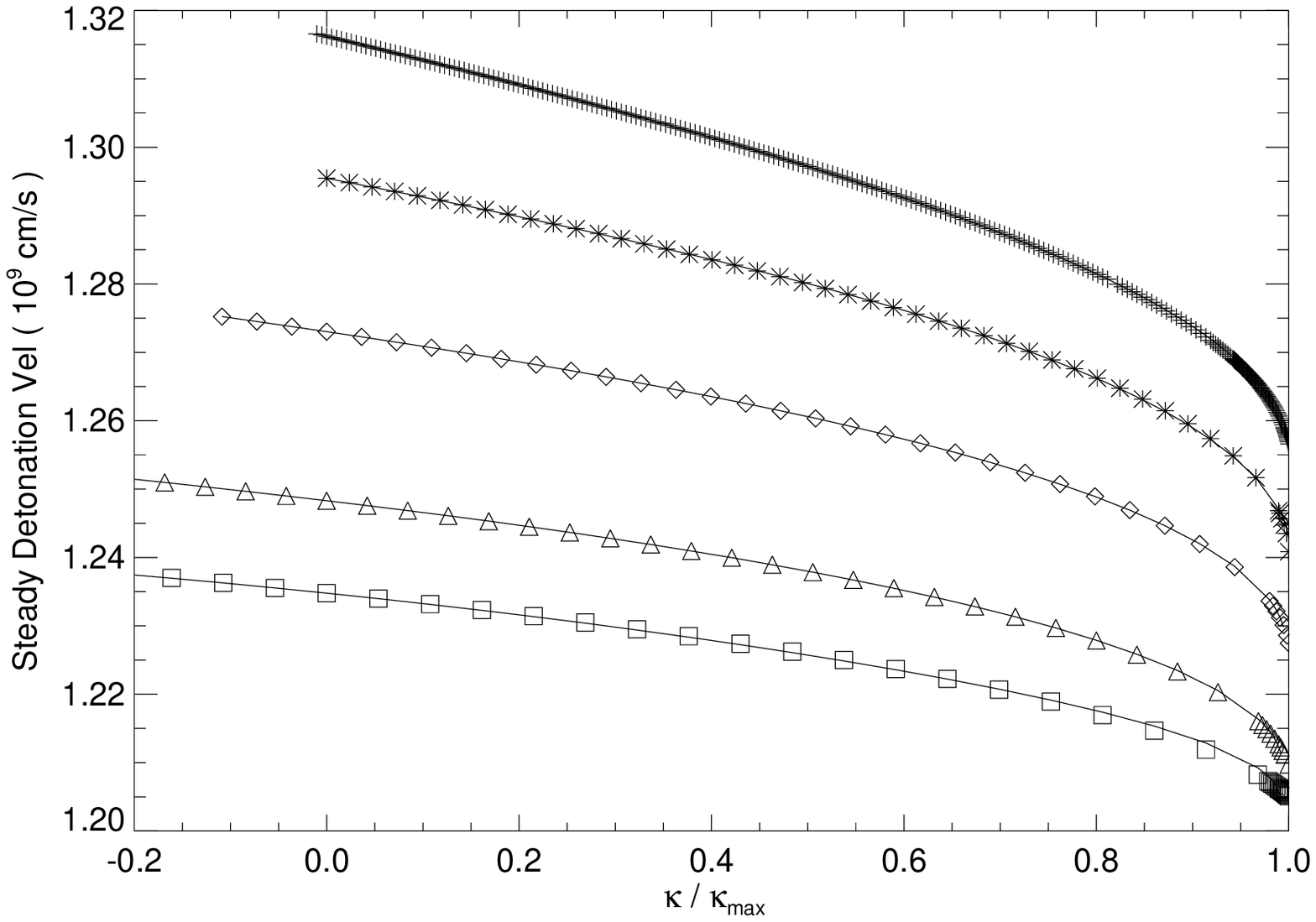}
\setlength{\columnwidth}{\curcolwidth}
\plottwo{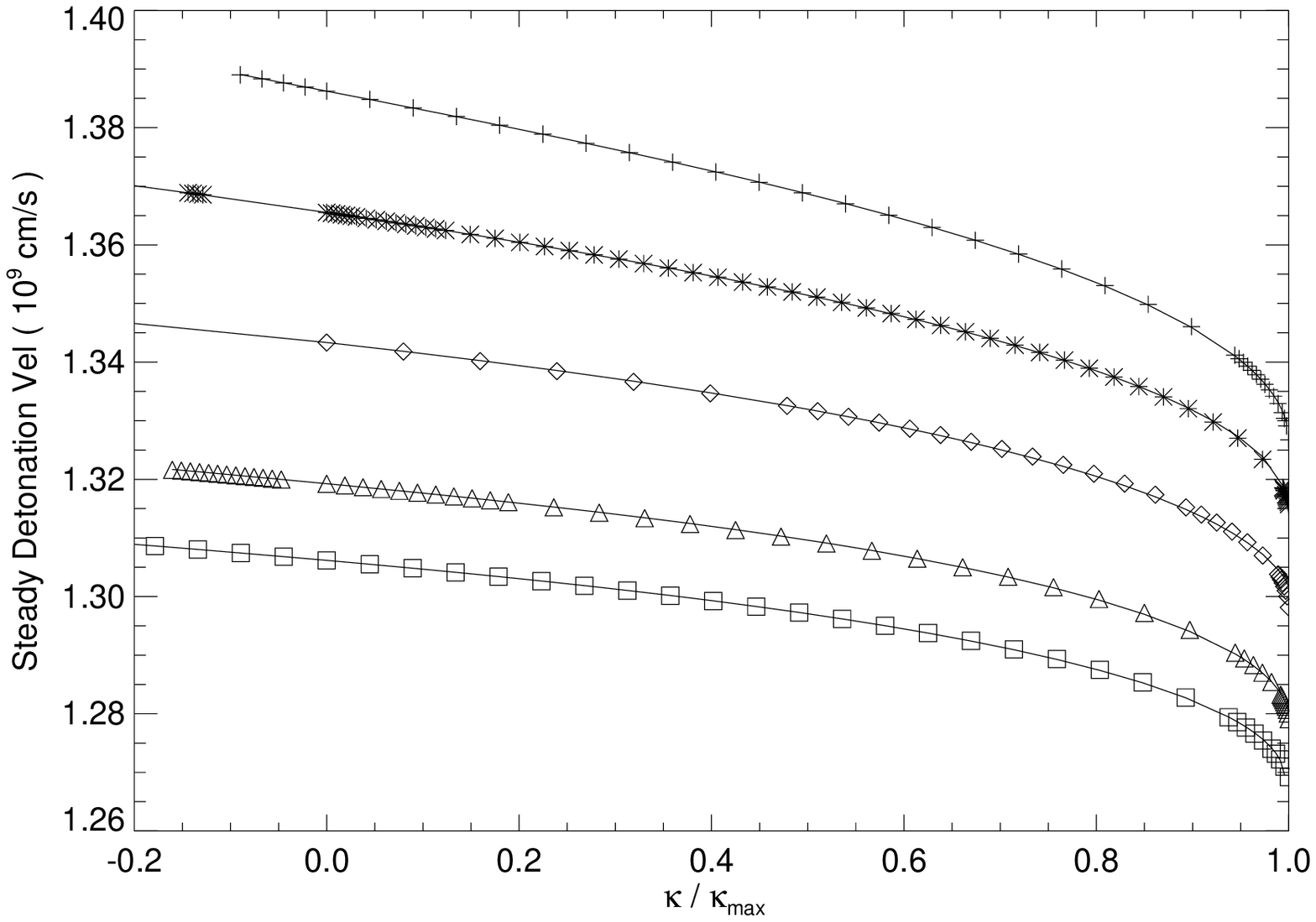}{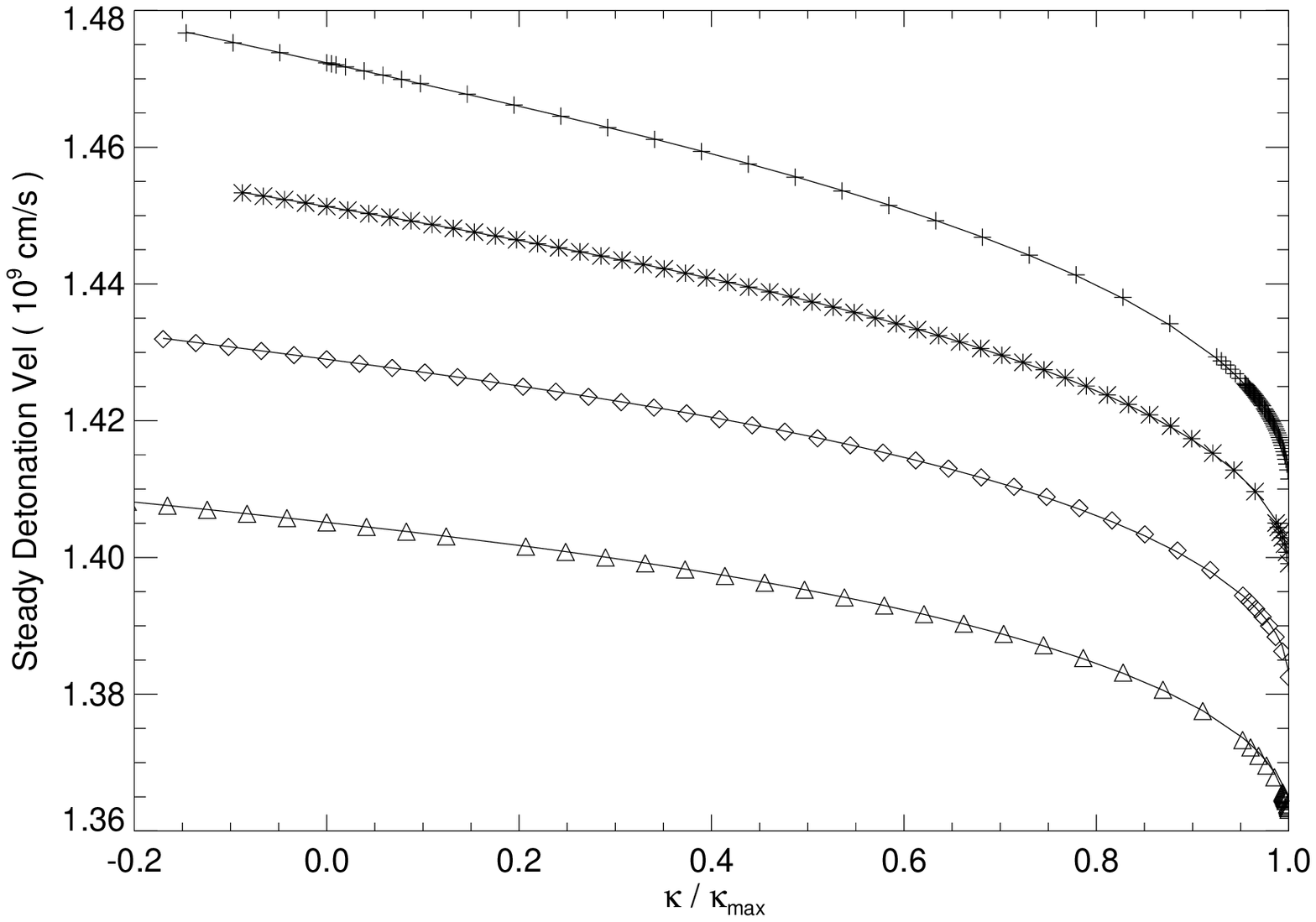}
\caption{Steady-state detonation velocity as a function of curvature
for background densities of, from left to right and top to bottom,
$\rho_8 = (0.5, 1, 2, 5, 10, 20)$.  In each plot, the velocities are
shown for initial carbon abundances, top to bottom, of
$\cfrac = (1, 0.75, 0.5, 0.25)$.  So that they could be
shown on the same horizontal scale, the curvatures have been scaled to the maximum
sustainable curvature for each set of conditions, given by
Table~\ref{table:detvelparams}.  Points represent calculated speeds,
and solid lines are fits with parameters also given in
Table~\ref{table:detvelparams}.}
\label{fig:detcurvevel}
\end{figure}

We calculated the detonation speed versus curvature relation for
$0.5 \le \rho_8 \le 20$ and $0.25 \le \cfrac \le 1.0$.  The unshocked
material was set to a temperature of $T_9 = 0.05$, although the results
were seen to be insensitive to this value.   Our results are shown
in Fig.~\ref{fig:detcurvevel} and Table~\ref{table:detvelparams}.
The code used to perform the calculations is available at
{\url{http://www.cita.utoronto.ca/$\sim$ljdursi/ignition}}.  The
estimated detonation thickness and a comparison with $\kmax$ is given
in Table~\ref{table:detthick}.   As compared with the \citet{heclavin}
results of $R_D \gtrsim 1000-1500 l_i$, we find $R_D \gtrsim 3000-6000 l_i$.

\begin{deluxetable}{lcccrrrr}
\tablecaption{Measured and Fit Detonation Velocities as a function of curvature}
\tablecolumns{9}
\tablewidth{0pt}
\tabletypesize{\footnotesize}
\tablehead{& \colhead{$D$\tablenotemark{a}} &  
             \colhead{$\kmax$\tablenotemark{b}} & 
             \colhead{$v_{\mathrm{max}}$\tablenotemark{c}} & \multicolumn{4}{c}{Fit Parameters} \\
           & \colhead{$(10^9 \cm)$} &  \colhead{$\cm^{-1}$} & \colhead{$10^9 \cm$} & \colhead{$a$\tablenotemark{d}} & \colhead{$b$} & \colhead{$c$} & \colhead{$d$} }
\startdata
$\rho_8 = 0.5$ \\
\tableline
$\cfrac = 1.00$ & $1.207$ & $1.08 \times 10^{-2}$ & $1.096$ & $-1.55 \times 10^{-2}$ & $7.75 \times 10^{-1}$ & $-1.26 \times 10^{0}$ &  $ 1.51 \times 10^{0}$  \\
$\cfrac = 0.75$ & $1.181$ & $2.94 \times 10^{-3}$ & $1.069$ & $-5.74 \times 10^{-4}$ & $4.68 \times 10^{-1}$ & $ 4.07 \times 10^{-1}$ & $ 1.26 \times 10^{-1}$ \\
$\cfrac = 0.50$ & $1.152$ & $7.14 \times 10^{-4}$ & $1.067$ & $-1.78 \times 10^{-3}$ & $6.11 \times 10^{-1}$ & $ 7.21 \times 10^{-1}$ & $-3.30 \times 10^{-1}$ \\
$\cfrac = 0.25$ & $1.120$ & $1.26 \times 10^{-4}$ & $1.055$ & $-2.27 \times 10^{-3}$ & $1.15 \times 10^{0}$ &  $-1.39 \times 10^{-1}$ & $-7.54 \times 10^{-3}$ \\
$\cfrac = 0.125$& $1.110$ & $2.34 \times 10^{-5}$ & $1.059$ & $-1.32 \times 10^{-2}$ & $1.20 \times 10^{0}$ &  $-1.65 \times 10^{-0}$ & $ 1.37 \times 10^{-3}$ \\
\tableline
$\rho_8 = 1.$ \\
\tableline
$\cfrac = 1.00$ & $1.218$ & $2.78 \times 10^{-2}$ & $1.131$ & $-4.69 \times 10^{-3}$ & $6.25 \times 10^{-1}$ &$ 1.87 \times 10^{-3}$ & $ 3.77 \times 10^{-1} $ \\
$\cfrac = 0.75$ & $1.195$ & $1.02 \times 10^{-2}$ & $1.116$ & $-4.08 \times 10^{-3}$ & $6.53 \times 10^{-1}$ &$ 3.30 \times 10^{-1}$ & $ 2.06 \times 10^{-2} $ \\
$\cfrac = 0.50$ & $1.170$ & $3.15 \times 10^{-3}$ & $1.102$ & $-5.41 \times 10^{-3}$ & $8.30 \times 10^{-1}$ &$ 2.64 \times 10^{-1}$ & $-8.85 \times 10^{-2} $ \\
$\cfrac = 0.25$ & $1.141$ & $5.58 \times 10^{-4}$ & $1.085$ & $-2.99 \times 10^{-3}$ & $1.17 \times 10^{0}$  &$-2.12 \times 10^{-1}$ & $ 4.44 \times 10^{-2} $ \\
$\cfrac = 0.125$& $1.126$ & $8.90 \times 10^{-5}$ & $1.092$ & $-9.94 \times 10^{-3}$ & $9.55 \times 10^{-1}$ &$ 1.35 \times 10^{-1}$ & $-7.99 \times 10^{-2} $ \\
\tableline
$\rho_8 = 2.$ \\
\tableline
$\cfrac = 1.00$ & $1.249$ & $6.67 \times 10^{-2}$ & $1.176$ & $-5.52 \times 10^{-4}$ & $7.93 \times 10^{-1}$ &$ 8.76 \times 10^{-2}$ & $ 1.20 \times 10^{-1} $ \\
$\cfrac = 0.75$ & $1.227$ & $2.72 \times 10^{-2}$ & $1.169$ & $-2.72 \times 10^{-3}$ & $6.75 \times 10^{-1}$ &$ 4.44 \times 10^{-1}$ & $-1.16 \times 10^{-1} $ \\
$\cfrac = 0.50$ & $1.203$ & $8.90 \times 10^{-3}$ & $1.152$ & $-1.07 \times 10^{-2}$ & $9.50 \times 10^{-1}$ &$ 1.07 \times 10^{-1}$ & $-4.54 \times 10^{-2} $ \\
$\cfrac = 0.25$ & $1.177$ & $1.56 \times 10^{-3}$ & $1.133$ & $-8.23 \times 10^{-3}$ & $1.10 \times 10^{0} $ &$-9.81 \times 10^{-2}$ & $ 2.26 \times 10^{-3} $ \\
$\cfrac = 0.125$& $1.163$ & $2.71 \times 10^{-4}$ & $1.127$ & $-1.07 \times 10^{-3}$ & $1.18 \times 10^{0} $ &$-2.05 \times 10^{-1}$ & $ 3.07 \times 10^{-2} $ \\
\tableline
$\rho_8 = 5.$ \\
\tableline
$\cfrac = 1.00$ & $1.316$ & $1.96 \times 10^{-1}$ & $1.257$ & $-5.81 \times 10^{-3}$ & $8.76 \times 10^{-1}$ &$ 1.20 \times 10^{-1}$ & $ 9.98 \times 10^{-3} $ \\
$\cfrac = 0.75$ & $1.295$ & $8.49 \times 10^{-2}$ & $1.241$ & $ 1.24 \times 10^{-3}$ & $1.11 \times 10^{0} $ &$-2.25 \times 10^{-1}$ & $ 1.11 \times 10^{-1} $ \\
$\cfrac = 0.50$ & $1.273$ & $2.76 \times 10^{-2}$ & $1.227$ & $-3.67 \times 10^{-3}$ & $1.16 \times 10^{0} $ &$-2.27 \times 10^{-1}$ & $ 6.94 \times 10^{-2} $ \\
$\cfrac = 0.25$ & $1.248$ & $4.75 \times 10^{-3}$ & $1.210$ & $-8.00 \times 10^{-3}$ & $1.19 \times 10^{0} $ &$-2.25 \times 10^{-1}$ & $ 4.18 \times 10^{-2} $ \\
$\cfrac = 0.125$& $1.235$ & $9.30 \times 10^{-4}$ & $1.205$ & $-9.26 \times 10^{-3}$ & $8.82 \times 10^{-1}$ &$ 2.83 \times 10^{-1}$ & $-1.56 \times 10^{-1} $ \\
\tableline
$\rho_8 = 10.$ \\
\tableline
$\cfrac = 1.00$ & $1.386$ & $4.45 \times 10^{-1}$ & $1.327$ & $-1.31 \times 10^{-4}$ & $1.04 \times 10^{0} $ &$-1.53 \times 10^{-1}$ & $ 1.10 \times 10^{-1} $ \\
$\cfrac = 0.75$ & $1.366$ & $1.94 \times 10^{-1}$ & $1.316$ & $-2.65 \times 10^{-3}$ & $1.08 \times 10^{0} $ &$-1.19 \times 10^{-1}$ & $ 4.04 \times 10^{-2} $ \\
$\cfrac = 0.50$ & $1.343$ & $6.27 \times 10^{-2}$ & $1.298$ & $ 6.40 \times 10^{-3}$ & $1.26 \times 10^{0} $ &$-3.40 \times 10^{-1}$ & $ 7.62 \times 10^{-2} $ \\
$\cfrac = 0.25$ & $1.319$ & $1.06 \times 10^{-2}$ & $1.279$ & $ 6.21 \times 10^{-4}$ & $1.40 \times 10^{0} $ &$-5.84 \times 10^{-1}$ & $ 1.81 \times 10^{-1} $ \\
$\cfrac = 0.125$& $1.306$ & $2.24 \times 10^{-3}$ & $1.269$ & $-5.93 \times 10^{-3}$ & $1.35 \times 10^{0} $ &$-4.74 \times 10^{-1}$ & $ 1.26 \times 10^{-1} $ \\
\tableline
$\rho_8 = 20.$ \\
\tableline
$\cfrac = 1.00$ & $1.472$ & $1.03 \times 10^{0}$ & $1.411$  & $ 1.03 \times 10^{-5}$ & $1.14 \times 10^{0} $ &$-2.99 \times 10^{-1}$ & $ 1.59 \times 10^{-1} $ \\
$\cfrac = 0.75$ & $1.451$ & $4.56 \times 10^{-1}$ & $1.399$ & $-1.93 \times 10^{-3}$ & $1.26 \times 10^{0} $ &$-4.25 \times 10^{-1}$ & $ 1.68 \times 10^{-1} $ \\
$\cfrac = 0.50$ & $1.429$ & $1.47 \times 10^{-1}$ & $1.382$ & $ 2.49 \times 10^{-4}$ & $1.41 \times 10^{0} $ &$-6.10 \times 10^{-1}$ & $ 2.03 \times 10^{-1} $ \\
$\cfrac = 0.25$ & $1.405$ & $2.41 \times 10^{-2}$ & $1.363$ & $-4.72 \times 10^{-3}$ & $1.43 \times 10^{0} $ &$-5.96 \times 10^{-1}$ & $ 1.68 \times 10^{-1} $ 
\enddata
\tablenotetext{a}{The measured planar ($\kappa = 0$) detonation speed.}
\tablenotetext{b}{The maximum curvature for which a steady-state self-sustained detonation can be found to persist.} 
\tablenotetext{c}{The detonation speed at the maximum curvature $\kmax$}
\tablenotetext{d}{The detonation velocity as a function of $\kappa$ can
be fit by $v' = \sqrt{a + b \kappa' + c {\kappa'}^2 + d{\kappa'}^3}$,
where $v' = (v - v_{\mathrm{max}})/(D - v_{\mathrm{max}})$ and $\kappa' =
(1 - \kappa/\kmax)$.   Note that for small curvature $\kappa$, one can
define a `Markstein length' \citep[\eg{},][]{flamecurve} of sorts by $d
v/d \kappa = -((D-v_{\mathrm{max}})/(2 \kmax)) (b + 2 c + 3 d)/(a + b +
c + d)^{1/2}$.}
\label{table:detvelparams}
\end{deluxetable}

\begin{deluxetable}{lrr}
\tablecaption{Detonation Thicknesses compared with $\kmax$}
\tablecolumns{3}
\tablewidth{0pt}
\tabletypesize{\footnotesize}
\tablehead{ & \colhead{$l_i$\tablenotemark{a}} & \colhead{$\kmax^{-1}/l_i$\tablenotemark{b}}}
\startdata
$\rho_8 = 0.5$ \\
\tableline
$\cfrac = 1.00$ & $2.26 \times 10^{-2}$ & $4.09 \times 10^{3}$\\
$\cfrac = 0.75$ & $5.48 \times 10^{-2}$ & $6.20 \times 10^{3}$\\
$\cfrac = 0.50$ & $1.78 \times 10^{-1}$ & $7.85 \times 10^{3}$\\
$\cfrac = 0.25$ & $1.09 \times 10^{0} $ & $7.27 \times 10^{3}$\\
\tableline
$\rho_8 = 1.$ \\
\tableline
$\cfrac = 1.00$ & $9.59 \times 10^{-3}$ & $3.74 \times 10^{3}$\\
$\cfrac = 0.75$ & $2.31 \times 10^{-2}$ & $4.23 \times 10^{3}$\\
$\cfrac = 0.50$ & $7.40 \times 10^{-2}$ & $4.29 \times 10^{3}$\\
$\cfrac = 0.25$ & $4.55 \times 10^{-1}$ & $3.93 \times 10^{3}$\\
\tableline
$\rho_8 = 2.$ \\
\tableline
$\cfrac = 1.00$ & $3.94 \times 10^{-3}$ & $3.81 \times 10^{3}$\\
$\cfrac = 0.75$ & $9.68 \times 10^{-3}$ & $3.80 \times 10^{3}$\\
$\cfrac = 0.50$ & $3.16 \times 10^{-2}$ & $3.55 \times 10^{3}$\\
$\cfrac = 0.25$ & $1.92 \times 10^{-1}$ & $3.33 \times 10^{3}$\\
\tableline
$\rho_8 = 5.$ \\
\tableline
$\cfrac = 1.00$ & $1.17 \times 10^{-3}$ & $4.35 \times 10^{3}$\\
$\cfrac = 0.75$ & $2.93 \times 10^{-3}$ & $4.02 \times 10^{3}$\\
$\cfrac = 0.50$ & $9.58 \times 10^{-3}$ & $3.78 \times 10^{3}$\\
$\cfrac = 0.25$ & $6.02 \times 10^{-2}$ & $3.50 \times 10^{3}$\\
\tableline
$\rho_8 = 10.$ \\
\tableline
$\cfrac = 1.00$ & $4.51 \times 10^{-4}$ & $4.99 \times 10^{3}$\\
$\cfrac = 0.75$ & $1.12 \times 10^{-3}$ & $4.58 \times 10^{3}$\\
$\cfrac = 0.50$ & $3.89 \times 10^{-3}$ & $4.10 \times 10^{3}$\\
$\cfrac = 0.25$ & $2.47 \times 10^{-2}$ & $3.82 \times 10^{3}$\\
\tableline
$\rho_8 = 20.$ \\
\tableline
$\cfrac = 1.00$ & $1.71 \times 10^{-4}$ & $5.67 \times 10^{3}$\\
$\cfrac = 0.75$ & $4.48 \times 10^{-4}$ & $4.89 \times 10^{3}$\\
$\cfrac = 0.50$ & $1.55 \times 10^{-3}$ & $4.40 \times 10^{3}$\\
$\cfrac = 0.25$ & $1.01 \times 10^{-2}$ & $4.09 \times 10^{3}$\\
\enddata
\tablenotetext{a}{Detonation thickness, in \cm{}, estimated by $l_i = v_s \tau_i$, where $v_s$ is the velocity of the shocked material
                  immediately behind the leading shock and $\tau_i$ is the ignition time at the same position.}
\tablenotetext{b}{$\kmax$ is taken from Table~\ref{table:detvelparams}.}
\label{table:detthick}
\end{deluxetable}

\begin{figure}[htb]
\centering
\plotone{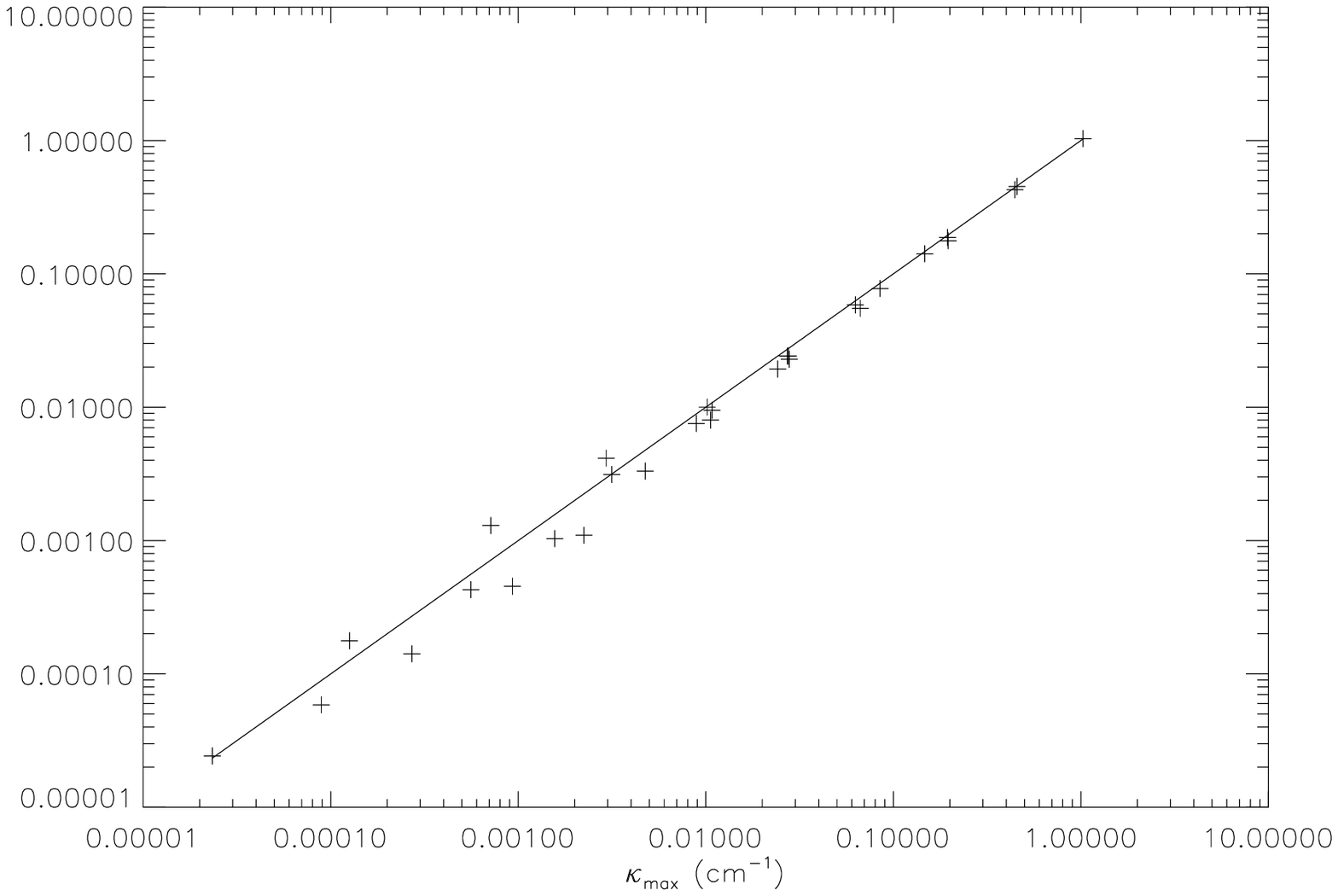}
\caption{Fit of maximum curvature for a sustainable steady-state detonation; the
fit is within 25\% for $\rho_8 \ge 1$, and within a factor of 2.5 for the entire
range.}
\label{fig:maxcurvefit}
\end{figure}

We can approximately summarize our results for the detonation velocity
and the maximum curvature of these curved detonations:
\begin{equation}
D = v(\kappa = 0) = 1.158 \times 10^9 \, \cms \left ( \frac{\cfrac}{0.5} \right )^{0.0273} \rho_8^{0.0678}
\end{equation}
\begin{equation}
\kmax = 3.128 \times 10^{-3} \, \cm^{-1} \left ( \frac{\cfrac}{0.5} \right )^{2.869} \rho_8^{1.272}
\label{eq:kmax}
\end{equation}
\begin{equation}
v_{\mathrm{max}} = v ( \kappa = \kmax )  = 1.098 \times 10^9 \, \cms \left ( \frac{\cfrac}{0.5} \right )^{0.0177} \rho_8^{0.0748}
\end{equation}
where $D$ is the planar detonation velocity, and $\kmax$
is the maximum sustainable curvature.  The fit for the maximum
curvature is shown in Fig.~\ref{fig:maxcurvefit}.  The velocity fits
produce errors of less than 10\% within the range examined; the fit
for $\kmax$ is better for highly degenerate material
(within 25\% for $\rho_8 \ge 1$) than for less-degenerate material
(within a factor of 2.5 for the entire range considered.)

One can estimate how much energy would have to be released at a point
to produce a shock wave that slowed to the detonation speed $D$ at a
radius of $R_D = \kmax^{-1}$ by using the similarity relations
of a Sedov blast wave propagating into a constant density material.
The position of the shock is given by \citep{sedov}:
\begin{equation}
r = \beta \left ( \frac{e}{\rho} \right )^{1/5} t^{2/5}
\end{equation}
where $\beta$ depends on the equation of state.  We performed numerical experiments for a Sedov blast
wave propagating through degenerate material at $\rho_8 = 1, T_9 = 0.05$ with no
burning and found $\beta \approx 0.77$.  Given this relation, the shock velocity would be
\begin{equation}
v_s = \frac{d r}{dt} = 4 \times 10^{11} \, \cms \beta^{5/2} \left ( \frac{e_{32}}{\rho_8} \right )^{1/2} r_0^{-3/2}
\end{equation}
where $e_{32}$ is the point energy input in units of
$10^{32}\,\mathrm{erg}$, and $r_0$ is the radius in units of $1 \,
\cm$.

Our numerical experiments for calculating the $\beta$ for the Sedov
blast wave under these conditions were as follows.  As described
later in this section for the calculation of spherical shock-ignited
detonations, we performed a series of one-dimensional spherically
symmetric hydrodynamical simulations, in this case without any burning,
for the hydrodynamic state described.   We used very high resolution,
with AMR ($\Delta x \approx 2 \times 10^{-4} \cm$, in a simulation box of
$5 - 50 \cm$ for these simulations without burning) to ensure that the
shock front was described adequately, and we placed the initial excess
energy uniformly in the center-most eight zones (\eg, over $1.5 \times
10^{-3} \cm$).   We found that this combination of high resolution and
finite size of energy input worked very well to produce the correct blast
wave structure.   We then tested an input of energy of $10^{26}, 10^{27},
10^{28}$, and $10^{29} \erg$, and for all input energies obtained a very
clean Sedov-like scaling between position (or velocity) and time, with
a scaling coefficient fit to be $0.77$.   As it turns out, these extra
calculations were likely unnecessary; the Sedov blast wave simulations
with burning also reproduce the same Sedov scaling with the same $\beta$
until the speed of the blast wave slows to within a factor two of the
planar detonation speed.

The requirement that the detonation be successful, that is $v_s \approx v_{\mathrm{max}}$
at $r \ge \kmax^{-1}$, gives a minimum energy
\begin{equation}
e_{32} \approx 246 \beta^{-5} \left ( \frac{\cfrac}{0.5} \right )^{-8.57} \rho_8^{-2.67}
\label{eq:minener}
\end{equation}
which for $\beta = 0.77$, $\rho_8 = 1$, $\cfrac = 1$ gives $e_{32} = 0.18$.

\begin{figure}[htb]
\centering
\plotone{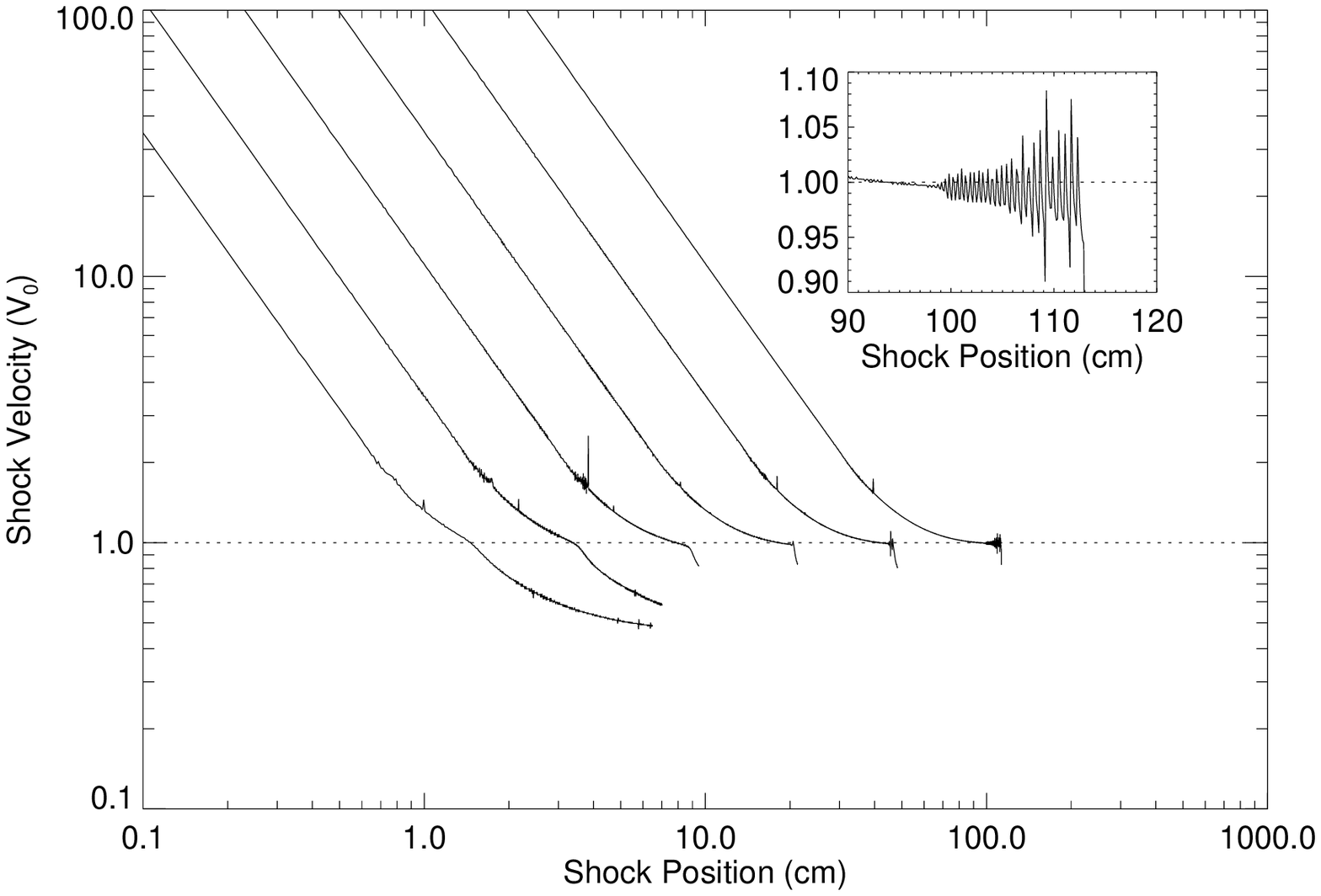}
\caption{ Shock velocity vs. shock position for a Sedov blast wave
with burning, propagating through constant density material.  The
velocity is measured in units of the steady planar detonation
velocity, $D = 1.218 \times 10^9$~cm/s.  Input energies are, left to
right, $10^{27}, 10^{28}, 10^{29}, 10^{30}, 10^{31}, 10^{32}$ ergs.
Note that when the shock velocity approaches the detonation velocity
from above, burning begins to effect the shock speed -- with transients
robustly occurring when the shock speed falls to about twice the planar
detonation velocity -- but the shock fails to sustain a detonation for the
case of input energies of less than $10^{31}$ ergs.  For the two largest
energies, a detonation does successfully propagate for many detonation
thicknesses -- which, for comparison purposes, is about $10^{-2} \cm$
-- but is disrupted by instabilities.  The inset shows a closeup of
the highest-energy `detonation' becoming unstable, although it propagates
$\approx 10^3 l_i$.}
\label{fig:detvel} 
\end{figure}

We can test the applicability of this energy criterion to detonations in
degenerate white dwarf material by running a reactive Sedov blast wave
through a constant density material with an given input point energy
and watch the success or failure of detonation ignition.  We performed
this calculation with the \FLASH{} code \citep{flashcode,flashvalid},
and the results are shown in Fig.~\ref{fig:detvel}.  The detonation was
resolved with a finest resolution of $\Delta x \approx 2 \times 10^{-4}
\cm$ (\eg{}, 60 points within $l_i \approx 1.2 \times 10^{-2} \cm$)
in a one-dimensional spherically symmetric domain of maximum radius
$512 \cm$.  A very high resolution was used to ensure the detonation
structure was adequately resolved; improperly resolving detonation can
greatly exaggerate curvature effects \citep{mbl96} as well as causing
spurious ignition or incorrect propagation speeds \citep{hydroandburning}.
At time zero, the energy initial energy that causes the blast wave was
deposited into a region of size $0.0015 \cm$, or about eight zones;
this is a large enough number of zones that the blast wave is well
resolved even at early times, but a small enough region that the Sedov
assumption of a point explosion remains valid, as demonstrated by the
Sedov scaling behavior until very late times when burning significantly alters
the shock solution.

As shown in Fig.~\ref{fig:detvel}, successful detonations do indeed
begin propagation at predicted the input energy, but they are disrupted
due to instabilities \citep[\eg{},][]{kbl98}; this suggests that the
energy requirement given here is a lower bound for successful detonation
ignition.  However, the results from these simulations need to be taken
schematically rather than quantitatively.  Because of the extremely
large point energies required to successfully ignite a detonation, for
much of the evolution shown in Fig.~\ref{fig:detvel}, the blast wave
is moving superluminally, as no relativistic effects were included in
the hydrodynamics!  Part of this is an artifact of assuming that the
energy input is at a very localized region.

\begin{figure}[htb]
\centering
\plotone{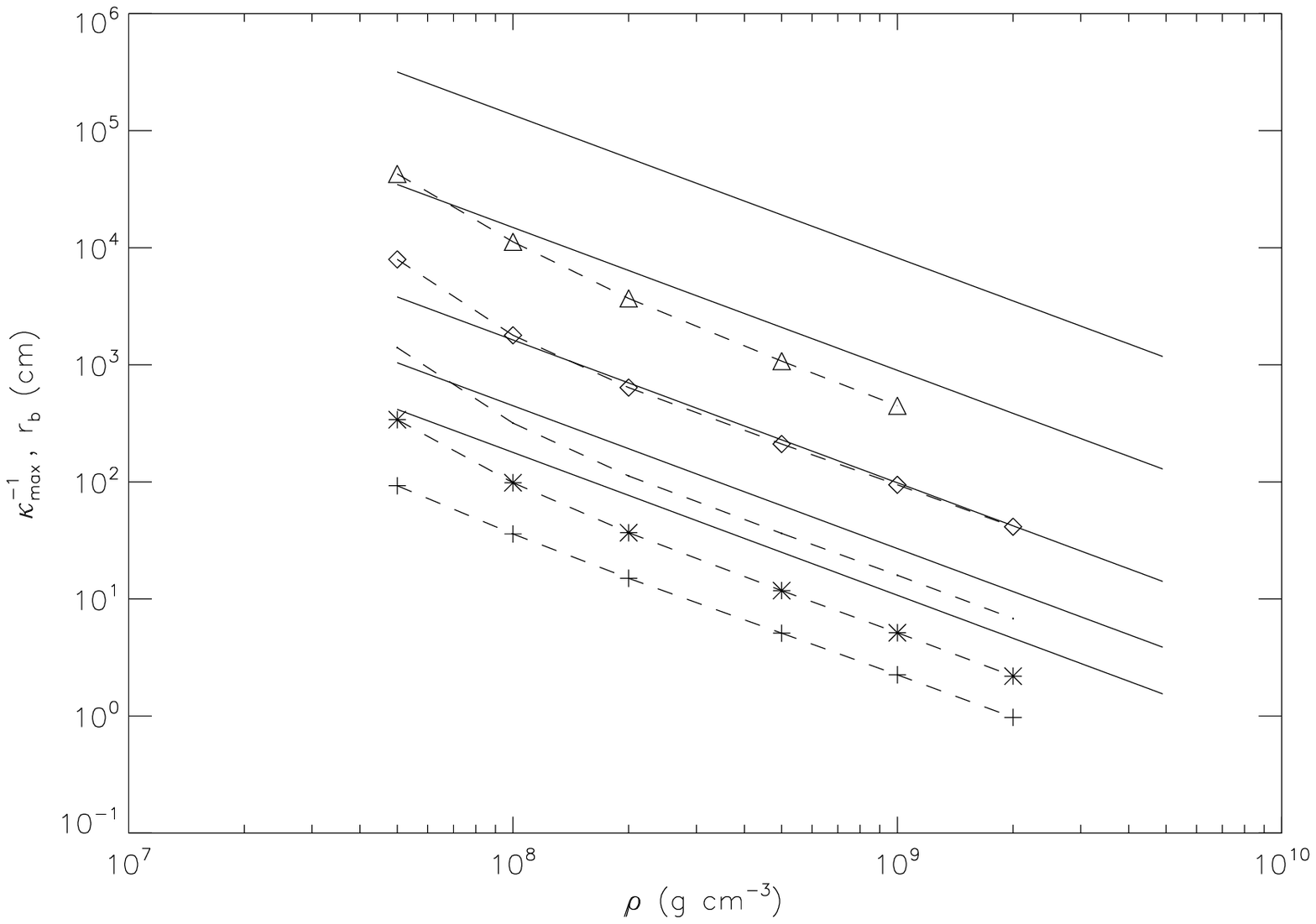}
\caption{Minimum scales for detonation.  Plotted is $\kmax^{-1}$ from
Table~\ref{table:detvelparams}, with dashed lines and points, and $r_b$
from Eq.~\ref{eq:rb} with solid lines.   An expending shock must slow
to approximately the steady detonation velocity at a radius no less
$\kmax^{-1}$, and $r_b$ is the radius in which is contained enough
potential nuclear energy from carbon burning to produce the energy for
a `Sedov' explosion which would would naturally slow down to $D$ at a
radius greater than $\kmax^{-1}$.   In both cases, lines are plotted for,
top to bottom, $\cfrac = 1/8, 1/4, 1/2, 3/4, 1$.   For the cases considered
here, $r_b > \kmax^{-1}$.}
\label{fig:sizes} 
\end{figure}

We can also consider the (more plausible) launch of a spherical
shock and possible ignition of detonation from a region of finite
size.  Constraints on what that size must be will come from energetic
considerations similar to that of the point explosion.  If the required
energy input comes predominantly from carbon burning
($\Delta \epsilon \approx 5.6\times 10^{17}~\mathrm{erg}~\mathrm{g}^{-1}$), 
then the volume of material which must be burned to provide such
energy can be found by setting the minimum energy requirement from
Eq.~\ref{eq:minener} to $4/3 \pi r_{b}^3 \rho \Delta \epsilon$, or, plugging in numerical
values, $2.09 \times 10^{-7} (10^{32} \mathrm{erg}) \rho_8 (\cfrac/0.5) r_{{b},0}^3$.
This gives a required radius of burning region 
\begin{equation}
r_{b} \approx 1.63 \times 10^{3} \cm \left ( \frac{\beta}{0.77} \right )^{-5/3} \left ( \frac{\cfrac}{0.5} \right ) ^ {-3.19} \rho_8^{-1.22}
\label{eq:rb}
\end{equation}
Note that in general this matchhead radius is actually larger than the
minimum radius for a sustainable detonation.   The two sizes are plotted
as a function of density in Fig.~\ref{fig:sizes}.

These results can be compared to empirical results from the astrophysics
literature, for instance \S~3.1.2 of \citet{niemeyerwoosley97}, where the
authors performed 100-zone spherically symmetric simulations to examine
the scales of detonation.   By imposing a linear temperature profile
peaking at $T_9 = 3.2$ over thirty zones of a given density and examining
the smallest that region could still ignite a detonation.  We compare our
results, in particular $\kmax^{-1}$, to the size of their matchheads.
Some caution should be exercised in this comparison, as they are of
different quantities --- in our work we find the smallest possible
radius of a steady-state detonation, whereas the quantity measured
in \citet{niemeyerwoosley97} is the smallest size of a region of a
particular imposed temperature profile which can launch a detonation.
With that caveat in mind, a comparison of the results where they
(nearly) overlap is given in Table~\ref{tab:nw-thisworkcompare}.
Because of the large amounts of energy imposed in the region by those
authors, and because the temperature profile imposed implies that the
detonation would normally `begin' where the temperature drops to that
of the ambient medium, the results are actually quite similar, with only the
one point at $T_9 = 2$ differing markedly.

\begin{deluxetable}{ccccc}
\tablecaption{Minimum scales for detonation ignition calculated here and in Niemeyer \& Woosley (1997)}
\tablecolumns{5}
\tablewidth{0pt}
\tabletypesize{\footnotesize}
\tablehead{ \colhead{$\rho_8$} & \colhead{\cfrac \tablenotemark{a}} & \colhead{\ofrac \tablenotemark{a}} & \colhead{NW97\tablenotemark{b}} & \colhead{this work\tablenotemark{c}} }
\startdata
1 & 1 & 0 & 40 \cm & 36 \cm \\
1 & $\frac{1}{2}$ & $\frac{1}{2}$ & 2 \metre & 3.17 \metre \\
20 & $\frac{1}{2}$ & $\frac{1}{2}$ & 70 \cm & 6.8 \cm \\
0.3 & 1 & 0 & 1 \metre & 30.2 \cm \tablenotemark{d} \\
0.3 & $\frac{1}{2}$ & $\frac{1}{2}$ & 50 \metre & 14.8 \metre \tablenotemark{d} \\
\enddata
\tablenotetext{a}{Mass fraction of Carbon-12 or Oxygen-16.}
\tablenotetext{b}{Size of region of imposed linear temperature profile, with peak temperature at $T_9 = 3.2$,
needed to provide a detonation in a 100-zone spherically symmetric simulation in \citet{niemeyerwoosley97}.}
\tablenotetext{c}{$\kmax^{-1}$ as measured in this section.}
\tablenotetext{d}{$\kmax^{-1}$ extrapolated from measurements using the fitting formula given in Eq.~\ref{eq:kmax}.}
\label{tab:nw-thisworkcompare}
\end{deluxetable}

\section{DISCUSSIONS AND CONCLUSIONS}

Any currently feasible mechanism for the explosion of a Type~Ia
supernova involves propagating a burning front in a carbon-oxygen
white dwarf.  Correctly modeling the process which leads to ignition is
crucial to understanding the resulting explosion; once a thermonuclear
burning front begins propagating it is very difficult to extinguish
\citep{zingaleflamevortex,niemeyerddt} and differing locations, numbers
and sizes of the ignition points can significantly change the large-scale
character of the explosion \citep[\eg{},][]{barcelona05,pcl}.

In this paper, we have examined ignition at a single point by calculating
and fitting ignition times over a range of thermodynamic conditions
and abundances relevant for almost any mechanism of ignition for a
Type~Ia supernova.  We have also considered the ignition of spherical
detonations, determining sizes necessary for a detonation successfully
ignite, and the energy required for a Sedov blast wave to lead to a
shock-ignited detonation.

Currently the favored model of SNIa ignition is for the central
density and temperature to increase as a carbon-oxygen white dwarf
grows by accretion from a main-sequence companion in a binary system.
When the energy deposited by carbon burning exceeds neutrino losses,
the white dwarf enters a simmering, convective stage that lasts
500-1000 years with convective velocities of order 20--100~\kms
\citep[\eg{},][]{hoeflichstein02,kuhlenwoosleyglatzmaier05}.  Owing to
the convection, the entropy of the core is almost constant.  As the
temperature continues to rise, thermonuclear flames are born at points
with the highest temperatures at or near the center of the white
dwarf.  Hot, Rayleigh-Taylor unstable bubbles begin to rise through
the convective interior.

In this turbulent convective picture of SNIa ignition, the first points
to runaway will be rare events at the high-temperature tail of the
turbulent distribution. Thus, it is important to accurately quantify
ignition conditions.  Our five-parameter fitting formulae
(equations \ref{eq:cpfit} and \ref{eq:cvfit}) reproduce our
computed ignition delay times very well over 15 orders of magnitude,
covering most thermodynamic conditions and compositions relevant for
ignition in a carbon-oxygen white dwarf.  In particular, we have found
that increases in the white dwarf metallicity, modeled by increases in
the \element{Ne}{20} abundance, can speed up the ignition delay time
by $\sim$ 30\% if the neon comes largely from oxygen, or decrease the
ignition delay time by a similar amount if it comes equally from carbon
and oxygen.  In the `rare event' picture, if all else is equal, this
suggests that the first ignition points may happen at significantly
different conditions in metal-poor white dwarfs than in metal-rich
progenitors -- either the first ignition points could occur when the
mean core temperature is somewhat lower, or more points could ignite.

\begin{figure}[htb]
\centering
\plottwo{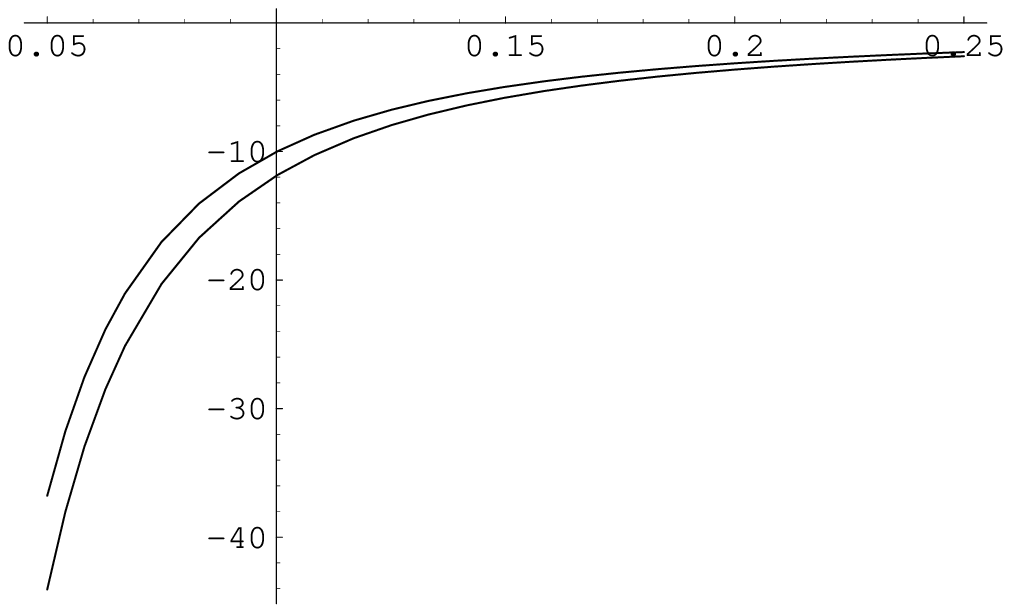}{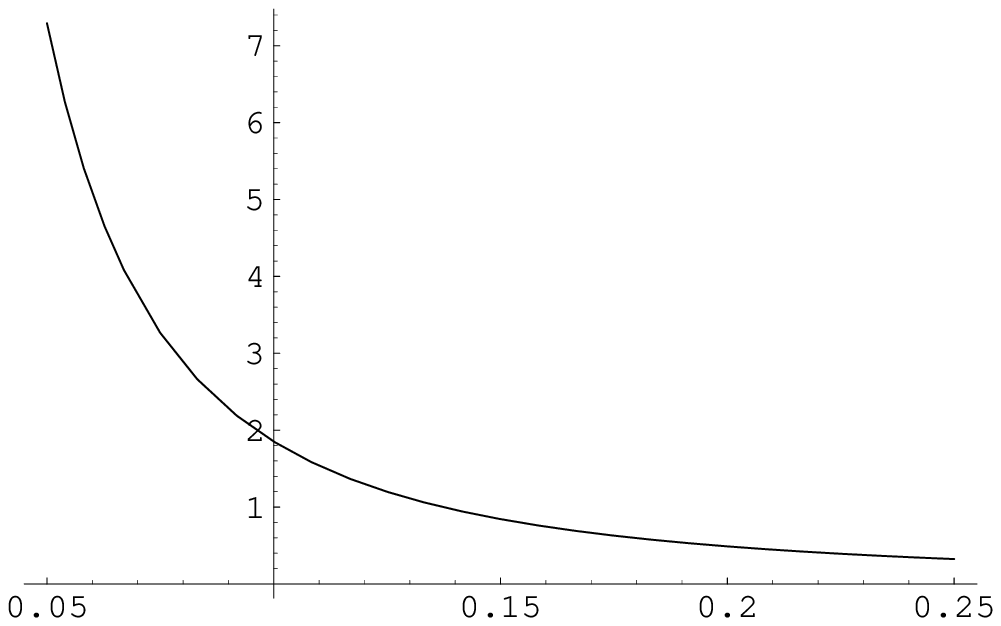}
\caption{Plots showing enhancement of probability of ignition for a 30\%
reduction in ignition time for a fiducial case ($T_9 = 2$, $\rho_8 =
10$, $\cfrac = 0.5$, and requiring ignition within $0.1 \mathrm{ns}$).
On the left is shown the base-10 logarithm of the probability of ignition
as a function of the RMS temperature fluctuation ($\delta T_{\mathrm{RMS}}/T$) with and without the 30\%
reduction in ignition time -- \eg{}, $\log_{10} P(\tau_{i,\mathrm{cp}} < 10^{-10} \sec)$ and $\log_{10} P(0.7 \tau_{i,\mathrm{cp}} < 10^{-10} \sec)$ --
for the constant pressure ignition time fit given in Eq.~\ref{eq:cpfit} and
assuming Gaussian distribution of temperature fluctuations.
On the right is shown the base-10 logarithm of the probability enhancement
of ignition with the 30\% reduction in ignition time, \eg{} $\log_{10}
[ P(0.7 \tau_{i,\mathrm{cp}} < 10^{-10} \sec) / P(\tau_{i,\mathrm{cp}}
< 10^{-10} \sec) ]$.  When ignition is very rare (such as $\delta
T_{\mathrm{RMS}} < 0.1 T$ for this case), the increase in probability
of ignition with a 30\% ignition temperature decrease can be substantial
(2--7 orders of magnitude).
}
\label{fig:probabilities} 
\end{figure}

To quantify this effect, Fig.~\ref{fig:probabilities} shows the
increase in the probability of a point igniting due to a 30\%
reduction in ignition time.  \citet{kuhlenwoosleyglatzmaier05} suggest
that the distribution of temperature fluctuations during the
simmering, convective phase is Gaussian, particularly at the
positive-fluctuation end.  Assuming this distribution and taking
$\delta T_{\mathrm{RMS}}$ as a free parameter that measures the
strength of turbulence, one finds that if ignition is extremely rare,
a 30\% reduction in ignition time could produce a substantial increase
in number in the (still rare) events.

Another example where local ignition can change burning behavior is
the propagation upwards through the star of a burning, Rayleigh-Taylor
unstable bubble.  Such a bubble will experience shear instabilities
and shed sparks \citep{mikebubble}.  Whether these sparks ignite the
unburned material into which they fall to produce more burning bubbles
(or fizzle out) will change the burning volume, and thus the overall
effectiveness of burning.  However, the statistics of these sparks are
not yet well characterized.

In the second part of our work, we examine another aspect of point
ignition -- the possibility of igniting a detonation at a point.
We have shown that very large amounts of energy are required to ignite
a detonation, making the purely local ignition of a detonation extremely
unlikely. In Fig.~\ref{fig:sizes}, we show the required minimum sizes for
the successful ignition of a steady state detonation.  The restriction on
the curvature of a successful detonation restricts all models for ignition
of a spherical detonation, for example placing limits on the minimum
size of a preconditioned region for successful ignition of a detonation
by the Zel'dovich gradient mechanism \citep[\eg{},][]{khokhlovddt}.

In particular, one oft-mooted possibility for the transition to a
detonation is for it to occur in the distributed burning regime near
$\rho_8 = 0.5$.  Multi-dimensional resolved calculations of burning in
this regime \citep{zingale2drtflames} has found that, because of both
the distributed burning itself and the resulting vigorous mixing of
fuel and ash, the remaining pockets of unburned fuel can have \cfrac as
low as 0.1--0.2.  Examining the case of $\cfrac = 0.125$ and $\rho_8 =
0.5$ we find that the restrictions on igniting a successful spherical
detonation are especially stringent, requiring matchheads on order half
a kilometer in size.  On the face of it, this makes preconditioning of an
ignition time gradient for the Zel'dovich mechanism seem fairly unlikely.

In our consideration of one-zone ignitions, we have ignored the influence
of hydrodynamics, which can transport energy into or away from a burning
region.   Further, in both the one-zone ignitions and our consideration
of spherical detonations, we have ignored thermal transport which, in the
electron-degenerate material of a white dwarf, can be rapid and efficient.
It is certainly the case that in regions which do successfully ignite,
the ignition timescale will dominate that of hydrodynamics or thermal
conductivity; however, without comparing this timescale to the competing
timescales of thermal and hydrodynamic transport, we have no way of
determining what the successfully igniting regions will be, and which
regions will `fizzle out'.   Thus the results presented here can give,
at best, one one piece of the story in determining ignition conditions.

Comparisons between the ignition times (or shock-crossing time of the
matchhead in the detonation case) to the hydrodynamic or thermal diffusion
ignition times is relatively straightforward if a back of the envelope
calculation is sufficient.   \citet{zingale3drtflames} indicates that
the buoyancy-driven turbulence relevant for ignition in SNIa follows a
Kolmogorov turbulence, in which case the hydrodynamic time for destruction
of a hot spot behaves as:
\begin{equation}
\tau_{\mathrm{hydro}}(l) \approx 4.3 \times 10^{-5} \sec 
\left (\frac{50 \kms}{V_{\mathrm{conv}}}\right ) 
\left (\frac{L_{\mathrm{conv}}}{100 \km}\right )^{1/3}
\left (\frac{l}{1 {\mathrm{m}}} \right )^{2/3}.
\end{equation}
where $V_{\mathrm{conv}}$ and $L_{\mathrm{conv}}$ are the velocity 
and length scales of the large-scale convective motions that provide
the stirring for the turbulence on the integral scales. 

Similarly, \citet{woosley04} gives approximate values for the
electron-conduction-dominated thermal diffusivity relevant for the core
of a white dwarf, for which one can find the diffusion time of a strong
(several times the background temperature) Gaussian hotspot of size $l$:
\begin{equation}
\tau_{\mathrm{diff}}(l) \approx 24 \sec \left ( \frac{l}{1 \metre} \right ).
\end{equation}

If one is in a regime when one of these timescales clearly `wins', then
this simple comparison of timescales is sufficient.   However, in the
case of the first ignition in the centre of a white dwarf, this will
not be the case, as these timescales will necessarily be quite close.
In this case, to understand the highly-nonlinear process of ignition,
one must rely on computational experiments with all of the relevant
physics included.   In future work, we will consider the ignition of
Gaussian hotspots in a quiescent medium with hydrodynamics and thermal
diffusion, and then consider the addition of another piece of physics --
the possibility of ignition under compression for the specific case of
a pulsational delayed detonation.

\bigskip 

\begin{acknowledgments}

LJD acknowledges the support of the National Science and Engineering
Research Council during this work.  FXT was supported by a National
Security Fellow position at Los Alamos National Laboratory.   The authors
thank M. Zingale for many comments on this manuscript, and in particular
suggesting that we consider $\cfrac = 0.125$ and the distributed burning
regime.  The authors also thank A. Calder for a close reading of this
text, A.~Karakas for useful discussion, and the anonymous referee
whose feedback improved this paper.  This work made use of NASA's
Astrophysical Data System.  The \FLASH{} software used for some of
this work was in part developed by the DOE-supported ASC / Alliance
Center for Astrophysical Thermonuclear Flashes at the University
of Chicago.  Other software used in this work is available at
{\url{http://www.cita.utoronto.ca/$\sim$ljdursi/ignition}} and
{\url{http://www.cococubed.com/code\_pages/codes.shtml}}.
Comments on this work are welcome at at
{\url{http://www.cita.utoronto.ca/$\sim$ljdursi/thisweek/}}.
\end{acknowledgments}

\clearpage
\appendix
\section{Constant-Pressure and Constant-Volume Burning}
\label{sec:thermburncoupling}

In this appendix, we describe the coupling of temperature evolution
and burning used in the reaction networks used in this work.

For a Helmholtz free energy based thermodynamic system, the temperature
$T$, density $\rho$, and molar composition vector $\bf Y$ are the
primary variables.  ($Y_i$ are related to the mass fractions $X_i$
used elsewhere in this paper through $X_i = A_i Y_i$, where $A_i$ is
the atomic mass number of the $i^{\mathrm{th}}$ element.)  We'll want to
expand thermodynamic quantities in terms of these variables, for  
example,
\begin{equation}
\label {eq1}
dP = \partf{P}{T}dT + \partf{P}{\rho}d\rho + \sum_i \partf{P}{Y_i}dY_i
\enskip.
\end{equation}

All reasonable stellar equations of state will return the $\partial
P/\partial T$ and $\partial P/\partial \rho$ partial derivatives
\citep[for a survey, see][]{timmesarnett99}. The last term in equation
(\ref{eq1}) is a bit tricker. Nearly all stellar equations of state use
the mean atomic weight ${\bar {\rm A}}$ and the mean charge ${\bar {\rm
Z}}$ to characterize the composition.  Thus, the last term in equation
(\ref{eq1}) can be expanded as
\begin{equation}
\label {eq2}
\sum_i \partf{P}{Y_i}dY_i =
\sum_i \partf{P}{{\bar{\rm A}}}  \partf{{\bar{\rm A}}}{Y_i}  dY_i
+
\sum_i \partf{P}{{\bar{\rm Z}}}  \partf{{\bar{\rm Z}}}{Y_i}  dY_i
\enskip .
\end{equation}
All complete stellar equations of state will return the $\partial
P/\partial {\bar {\rm A}}$ and $\partial P/\partial {\bar {\rm Z}}$
terms, so we will not be concerned with those terms
\citep[\eg{},][]{eos}.  In terms of the molar composition the mean  
atomic
weight ${\bar {\rm A}}$ and the mean charge ${\bar {\rm Z}}$ are
\begin{equation}
\label {eq3}
{\bar{\rm A}}  = {1 \over \sum_i Y_i}
\quad \quad
{\bar{\rm Z}}  = {\sum_i Z_i Y_i \over \sum_i Y_i}
\enskip.
\end{equation}
The partial derivatives of ${\bar {\rm A}}$ and ${\bar{\rm Z}}$ are then
\begin{equation}
\label {eq4}
\partf{{\bar{\rm A}}}{Y_i}  = - \left ( {1 \over \sum_i Y_i} \right )^
{2}
= -{\bar{\rm A}}^2
\end{equation}
\begin{eqnarray}
\label {eq5}
\partf{{\bar{\rm Z}}}{Y_i}  & = &
-\left ( {1 \over \sum_i Y_i} \right )^{2} \sum_i Z_i Y_i
+ {Z_i  \over \sum_i Y_i} \nonumber \\
& = &
-{\bar{\rm A}}^2 \sum_i Z_i Y_i
+ {\bar{\rm A}}^2 Z_i \nonumber  \\
& = &
{\bar{\rm A}}
\left ( Z_i - {\bar{\rm A}} \sum_i Z_i Y_i \right ) \nonumber \\
& = &
{\bar{\rm A}}
\left ( Z_i - {\bar{\rm Z}}  \right )
\end{eqnarray}
surprisingly simple. Operating with $d/dt$ on equation (\ref{eq1})
yields
a pressure evolution equation
\begin{equation}
\label {eq6}
\drvf{P}{t} = \partf{P}{T}{dT\over dt} + \partf{P}{\rho} \drvf{\rho}{t}
+
\partf{P}{{\bar{\rm A}}} \sum_i \partf{{\bar{\rm A}}}{Y_i} \drvf{Y_i}{t}
+
\partf{P}{{\bar{\rm Z}}} \sum_i \partf{{\bar{\rm Z}}}{Y_i}\drvf{Y_i}{t}
\end{equation}
The last two terms of equation (\ref{eq6}) are easy to compute since
the $dY_i/dt$ are the right-hand sides of the ordinary differential
equations that comprise a nuclear reaction network equation.

The First Law of Thermodynamics
\begin{equation}
\label {eq7}
dE + P d \left( {1 \over \rho} \right ) = 0
\enskip ,
\end{equation}
operated on with $d/dt$ in the presence of an energy source becomes
\begin{equation}
\label {eq8}
\drvf{E}{t} - {P \over \rho^2} \drvf{\rho}{t} = {\dot S}
\enskip .
\end{equation}
Expanding the specific internal energy $dE/dt$ term
\begin{equation}
\label {eq9}
\partf{E}{T}\drvf{T}{t} + \partf{E}{\rho} \drvf{\rho}{t}
+
\partf{E}{{\bar{\rm A}}} \sum_i \partf{{\bar{\rm A}}}{Y_i} \drvf{Y_i}{t}
+
\partf{E}{{\bar{\rm Z}}} \sum_i \partf{{\bar{\rm Z}}}{Y_i}\drvf{Y_i}{t}
- {P \over \rho^2} \drvf{\rho}{t}
= {\dot S}
\end{equation}
and collecting like terms becomes
\begin{equation}
\label {eq10}
\partf{E}{T}\drvf{T}{t}
+
\left [\partf{E}{\rho} - {P \over \rho^2} \right ] \drvf{\rho}{t}
+
\partf{E}{{\bar{\rm A}}} \sum_i \partf{{\bar{\rm A}}}{Y_i} \drvf{Y_i}{t}
+
\partf{E}{{\bar{\rm Z}}} \sum_i \partf{{\bar{\rm Z}}}{Y_i}\drvf{Y_i}{t}
= {\dot S}
\end{equation}

For a constant density evolution, $d\rho = 0$, and equation (\ref{eq10})
may be written as
\begin{equation}
\label {eq11}
\drvf{T}{t}  =
{1 \over \partf{E}{T}}
\left [
{\dot S}
-
\partf{E}{{\bar{\rm A}}} \sum_i \partf{{\bar{\rm A}}}{Y_i} \drvf{Y_i}{t}
-
\partf{E}{{\bar{\rm Z}}} \sum_i \partf{{\bar{\rm Z}}}{Y_i}\drvf{Y_i}{t}
\right ]
\enskip .
\end{equation}
Equation (\ref{eq11}) obeys the First Law of Thermodynamics while being
consistent with a general stellar equation of state and the constraint
of constant density (no PdV work). To self-consistently couple the  
thermodynamic and
abundance evolutions we implicitly solve equation (\ref{eq11}) with  
the constraint
of a constant density and the reaction network equations
\begin{eqnarray}
\label {eq12}
\drvf{\rho}{t}  & = & 0 \\
\drvf{Y_i}{t}  & = & \sum_j C_i R_j Y_j
             + \sum_{j,k} {C_i \over C_j! C_k! } \rho N_A R_{j,k} Y_j  
Y_k
             + \sum_{j,k,l} {C_i \over C_j! C_k! C)l!} \rho^2 N_A^2 R_ 
{j,k} Y_j Y_k Y_l
\end{eqnarray}
as a single, fully coupled system of ordinary differential
equations.  We use the semi-implicit Bader-Deflhard algorithm with
the MA28 sparse matrix package as described in \citet{burn}. to solve
equations (\ref{eq11}) and (\ref{eq12}) simultaneously.  In equation
(\ref{eq12}) the (stoichiometric) coefficients $C_i$ can be positive
or negative numbers that specify how many particles of species $i$ are
created or destroyed. The factorials in the denominator avoid double
counting. The $R_i$ are the reaction rate for the species involved,
with the first sum representing reactions involving a single nucleus, the
second sum representing binary reactions, and the third sum representing
three-particle processes.

For a constant pressure evolution, $dP = 0$, equation (\ref{eq6})  
becomes
\begin{equation}
\label {eq13}
\partf{P}{T} \drvf{T}{t} + \partf{P}{\rho} \drvf{\rho}{t}
+
\partf{P}{{\bar{\rm A}}} \sum_i \partf{{\bar{\rm A}}}{Y_i} \drvf{Y_i}{t}
+
\partf{P}{{\bar{\rm Z}}} \sum_i \partf{{\bar{\rm Z}}}{Y_i}\drvf{Y_i}{t}
= 0
\enskip .
\end{equation}
Solving for $d\rho/dt$ yields
\begin{equation}
\label {eq14}
\drvf{\rho}{t} =
- {1 \over \partf{P}{\rho}}
\left [
\partf{P}{T}\drvf{T}{t}
+
\partf{P}{{\bar{\rm A}}} \sum_i \partf{{\bar{\rm A}}}{Y_i} \drvf{Y_i}{t}
+
\partf{P}{{\bar{\rm Z}}} \sum_i \partf{{\bar{\rm Z}}}{Y_i}\drvf{Y_i}{t}
\right ]
\enskip .
\end{equation}
Substituting equation (\ref{eq14}) into equation (\ref{eq10})
\begin{eqnarray}
\label {eq15}
\partf{E}{T}\drvf{T}{dt}
& - &
\left [ {\partf{E}{\rho} - {P \over \rho^2} \over \partf{P}{\rho}}
\right ]
\left [
\partf{P}{T}\drvf{T}{t}
+
\partf{P}{{\bar{\rm A}}} \sum_i \partf{{\bar{\rm A}}}{Y_i} \drvf{Y_i}{t}
+
\partf{P}{{\bar{\rm Z}}} \sum_i \partf{{\bar{\rm Z}}}{Y_i}\drvf{Y_i}{t}
\right ] \\
& + &
\partf{E}{{\bar{\rm A}}} \sum_i \partf{{\bar{\rm A}}}{Y_i} \drvf{Y_i}{t}
+
\partf{E}{{\bar{\rm Z}}} \sum_i \partf{{\bar{\rm Z}}}{Y_i}\drvf{Y_i}{t}
= {\dot S} \nonumber
\end{eqnarray}
and collecting like terms yields
\begin{eqnarray}
\label {eq16}
& & \left [
\partf{E}{T} -
\left ( {\partf{E}{\rho} - {P \over \rho^2} \over \partf{P}{\rho}}
\right )
\drvf{P}{T}
\right ]
\drvf{T}{t}  + \\
& & \left [
\partf{E}{A}  -
\left ( {\partf{E}{\rho} - {P \over \rho^2} \over \partf{P}{\rho}}
\right )
\drvf{P}{A}
\right ]
\sum_i \partf{{\bar{\rm A}}}{Y_i} \drvf{Y_i}{t} + \nonumber \\
& & \left [
\partf{E}{Z} -
\left ( {\partf{E}{\rho} - {P \over \rho^2} \over \partf{P}{\rho}}
\right )
\drvf{P}{Z}
\right ]
\sum_i \partf{{\bar{\rm Z}}}{Y_i} \drvf{Y_i}{t} = {\dot S} \nonumber
\end{eqnarray}
and solving for $dT/dt$ yields
\begin{eqnarray}
\label {eq17}
\drvf{T}{t}  & = &
\left [
\partf{E}{T}
- \left ( {\partf{E}{\rho} - {P \over \rho^2} \over \partf{P}{\rho}}
\right )
\drvf{P}{T}
\right ]^{-1} \times \\
& &
\left [  {\dot S}
- \left [
\partf{E}{A}
- \left ( {\partf{E}{\rho} - {P \over \rho^2} \over \partf{P}{\rho}}
\right )
\drvf{P}{A}
\right ]
\sum_i \partf{{\bar{\rm A}}}{Y_i} \drvf{Y_i}{t} -
\left [
\partf{E}{Z}
- \left ( {\partf{E}{\rho} - {P \over \rho^2} \over \partf{P}{\rho}}
\right )
\drvf{P}{Z}
\right ]
\sum_i \partf{{\bar{\rm Z}}}{Y_i} \drvf{Y_i}{t}
\right ] \nonumber
\enskip .
\end{eqnarray}
Equation (\ref{eq17}) obeys the First Law of Thermodynamics  
(including PdV
work) while being consistent with a general stellar equation of state
and the constraint of constant pressure.  To self-consistently couple
the thermodynamic and abundance evolutions we implicitly solve
equation (\ref{eq17}) along with the constraint of a constant pressure  
and the
reaction network equations
\begin{eqnarray}
\label {eq18}
\rho & = & f(P - P_{{\rm const}} = 0) \\
\drvf{Y_i}{t}  & = & \sum_j C_i R_j Y_j
             + \sum_{j,k} {C_i \over C_j! C_k! } \rho N_A R_{j,k} Y_j  
Y_k
             + \sum_{j,k,l} {C_i \over C_j! C_k! C_l!} \rho^2 N_A^2 R_ 
{j,k} Y_j Y_k Y_l
\end{eqnarray}
as a single, fully coupled system of differential-algebraic equations.
We use the same integration method and linear algebra solver listed
above to solve equations (\ref{eq17}) and (\ref{eq18}) simultaneously.
Given the constant upstream pressure, temperature, and composition,
a root-find is performed to find the density $\rho$ in equation
{\ref{eq18}} such that the pressure returned from an equation
of state is equal to the constant upstream pressure.  Note the
similarities between equation (\ref{eq11}) for constant density and
equation (\ref{eq17}) for constant pressure. Despite its complex
appearance, equation (\ref{eq17}) can be concisely coded.  The method
described here is implemented in the burning networks found at
{\url{http://www.cococubed.com/code\_pages/codes.shtml}}.

A complementary approach for isobaric combustion evolution was recently
formulated by \citet{2004ApJS..151..345C}.  Although their approach is
tied to the first-order Euler integration method, they demonstrated
satisfactory accuracy and consistency of the thermodynamic and abundance
evolutions. A direct comparison of the two methods should be examined
in future investigations.

\section{Calculation of Steady-State Detonation Structures}
\label{sec:steadystatedet}

Here we describe our method, which is almost completely as described in 
\citet{sharpecurved}, for calculating the steady state velocities
for unsupported detonations under curvature.

The starting point is to consider the quasi-steady equations of
hydrodynamics in shock attached frame, assuming curvature is constant
through structure.  This gives us Sharpe's eqns (6)--(9):

\begin{equation}
\frac{d \rho u_n}{dn} + \kappa \rho \left ( u_n + D_n \right ) = 0
\label{eq:steadycontinuity}
\end{equation}

\begin{equation}
u_n \frac{d u_n}{dn} + \frac{1}{\rho} \frac{d p}{d n} = 0
\label{eq:steadymomentum}
\end{equation}

\begin{equation}
\frac{d e}{dn} - \frac{p}{\rho^2} \frac{d \rho}{d n} = 0
\label{eq:steadyenergy}
\end{equation}

\begin{equation}
\frac{d X_i}{dn} = \frac{R_i}{u_n}
\label{eq:steadyabundance}
\end{equation}

where $n$ is the coordinate normal to the shock, $u_n$ and $D_n$ are the
fluid velocities relative to the shock and $D_n$ is the normal detonation
velocity, $X_i$ is the mass abundance of the $i^{\mathrm{th}}$ species,
$R_i$ is net production rate of the $i^{\mathrm{th}}$ species, $p$
is the gasdynamic pressure, and $e$ is total specific energy of the fluid.

Using arguments similar to those described in the previous appendix, one
can construct ODEs for the post-shock density and temperature structure
in the detonation:

\begin{equation}
\frac{d \rho}{d n} = \frac{\phi}{a_f^2 - u_n^2},
\label{eq:steadyrhostructure}
\end{equation}

\begin{equation}
\frac{d T}{d n} = \left ( \partf{p}{T} \right )_{\rho,{\bf{Y}}}^{-1} 
 \left \{ \left [ u_n^2 - \left ( \partf{p}{\rho} \right )_{T,{\bf{Y}}} \right ] \frac{d \rho}{d n} - \sum_{i=1}^{13} \left ( \partf{p}{X_i} \right )_{\rho,T,Y_j,j\ne i} \frac{d X_i}{d n} + \kappa \rho u_n ( u_n + D_n ) \right \}.
\label{eq:steadytempstructure}
\end{equation}

where $a_f$, the frozen sound speed, is given by 
\begin{equation}
a_f^2 = \left ( \partf{p}{\rho} \right )_{T,{\bf{Y}}} + 
        \left[ \frac{p}{\rho^2} - \left ( \partf{e}{\rho} \right )_{T,{\bf{Y}}} \right ]
        \left ( \partf{p}{T} \right )_{\rho,{\bf{Y}}} 
        \left ( \partf{e}{T} \right )_{\rho,{\bf{Y}}} ^{-1},
\label{eq:steadyfrozensoundspeed}
\end{equation}

and $\phi$, the thermicity, is given by 
\begin{equation}
\phi = \sum_{i=1}^{13} \left \{ \left ( \partf{p}{X_i} \right )_{\rho, T, Y_j,j\ne i}
 - \left [ \left ( \partf{e}{X_i} \right )_{\rho,T,Y_j,j\ne i} - \left ( \partf{q}{X_i} \right )_{Y_j, j\ne i} \right ] \left ( \partf{p}{T} \right )_{\rho,{\bf{Y}}} \left ( \partf{e}{T} \right)_{\rho,{\bf{Y}}}^{-1} \right \} \frac{d X_i}{d n} + \kappa \rho u_n ( u_n + D_n ).
\label{eq:thermicity}
\end{equation}

Note that at the sonic point -- as $u_n \rightarrow a_f$ -- the density
becomes unbounded unless at the same time, $\phi \rightarrow 0$.
This pair of conditions defines the generalized sonic point, and defines
a required boundary value at that singular point.

Normally it would be best to integrate from the singular point to the
shock that defines the detonation front, but we do not have enough initial
conditions to do this.  Instead we use a shooting method; given an input
range of detonation velocities, we pick the middle value and integrate
these equations using an IVODE solver until we hit the singular point.
If the initial guess was close enough to the correct detonation velocity,
then if we hit $\phi = 0$ before the sonicity becomes zero, our guessed
velocity was too high; if we hit $u_n = a_f$ first, the guessed velocity
was too low.  We then continue bisecting the range and choosing the new
midpoint a fixed number of times until we have the velocity within a
satisfactorily small range.  (The velocities were calculated to
ranges much smaller --- 2--20\cms --- than suggested  by the number of
significant figures quoted in our results here).

The only difference in how we proceeded compared to \citet{sharpecurved}
concerned how the partial derivatives were taken with respect to $X_i$.
Sharpe used a 13-isotope burning network, as do we, but did not evolve the
13th isotope, instead explicitly using the definition of mass fraction,
\begin{equation} \sum X_i = 1.  \end{equation}

Thus (presumably) differentiating by $X_i$ was performed computing finite
differences by varying $X_i$, and implicitly changing $X_{13}$ by the
same amount in the opposite direction.   However, there are arbitrarily
many other possibilities that could be used to `redistribute' the change
made to $X_i$.   We found it more accurate to consider the changes to
the other mass fractions to be proportional to the amount the abundances
were varying along the detonation structure at that point; these values
were available from the burning routines.   

The code used in computing these structures and velocities is available
at {\url{http://www.cita.utoronto.ca/$\sim$ljdursi/ignition}}.

%
%
\bibliographystyle{plainnat}
\bibliography{ignition}

\end{document}